\newif\ifonecol

\onecolfalse     

\ifonecol
\documentclass[11pt,draftcls,onecolumn]{IEEEtran}
\else
\documentclass[journal]{IEEEtran}
\fi
\IEEEoverridecommandlockouts
\usepackage{amsmath,epsfig,amssymb,amsthm,url}

\newcommand{\Rr}{\mathbb{R}}
\newcommand{\Prob}{\mathbb{P}}

\newcommand{\eb}{{\bf e}}
\newcommand{\Cb}{{\bf C}}
\newcommand{\nb}{{\bf n}}

\newcommand{\Ab}{{\bf A}}
\newcommand{\Abbar}{{\bf \bar A}}

\newcommand{\Bb}{{\bf B}}
\newcommand{\Bbbar}{{\bf B}^c}

\newcommand{\Xb}{{\bf X}}
\newcommand{\Yb}{{\bf Y}}

\newcommand{\Bpinv}{{\Bb^{\dag}}}

\newcommand{\ab}{{\bf a}}
\newcommand{\bb}{{\bf b}}
\newcommand{\xb}{{\bf x}}
\newcommand{\yb}{{\bf y}}

\newcommand{\sbb}{{\bf s}}
\newcommand{\hsbb}{{\bf \hat s}}

\newcommand{\deltab}{{\mbox{\boldmath $\delta$}}}

\newcommand{\Mb}{{\bf M}}
\newcommand{\Nb}{{\bf N}}

\newcommand{\ie}{{\em i.e.}}
\newcommand{\eg}{{\em e.g.}}

\newcommand{\argmin}{\mathop{{\rm argmin}}}
\newcommand{\spark}{\mbox{\sl{spark}}}

\newcommand{\Pc}{{\cal P}}

\newcommand{\sm}[2]{\sigma_{\textrm{\rm min}}^{(#1)}(#2)}
\newcommand{\varsm}[2]{\sigma_{\textrm{\rm min},#2}^{(#1)}}

\newcommand{\ga}{G_{\!\Ab}}
\newcommand{\nz}[1]{\|#1\|_0} 
\newcommand{\nt}[1]{\|#1\|_2} 
\newcommand{\la}{\left\{} 
\newcommand{\ra}{\right\}} 
\newcommand{\spx}{\sqrt{p-x}}
\newcommand{\bsx}{b-\sqrt{x}}

\newtheorem{defi}{Definition}
\newtheorem{theorem}{Theorem}
\newtheorem{lemma}{Lemma}
\newtheorem{prop}{Proposition}
\newtheorem{coro}{Corollary}

\begin{document}

\title{On the error of estimating the sparsest solution of underdetermined linear systems}

 \author{Massoud Babaie-Zadeh,~\IEEEmembership{Senior Member, IEEE,} 
   Christian Jutten,~\IEEEmembership{Fellow, IEEE,}
   Hosein Mohimani
  \thanks{This work has been partially funded by Iran Telecom Research Center (ITRC) and Iran National Science Foundation (INSF). Part of this work has been done while the first author was at sabbatical in Digital Technology Center (DTC), University of Minnesota, so he would like to thank them, too.}
  \thanks{Massoud Babaie-Zadeh is with Electrical Engineering Department, Sharif University of
    Technology, Tehran, Iran (e-mail: {\tt mbzadeh@yahoo.com}).}
  \thanks{Christian Jutten is with GIPSA-Lab, Grenoble, France, and is a member of Institut Universitaire de France (e-mail: {\tt Christian.Jutten@inpg.fr}).}
  \thanks{Hosein Mohimani is a Ph.D.~student at the Department of Electrical Engineering, University of California, San Diego, USA (e-mail: {\tt hmohiman@ucsd.edu}).}
  \ifonecol
  \thanks{Corresponding author: Massoud Babaie-Zadeh, email: {\tt mbzadeh@yahoo.com}, Tel: +98 21 66 16 59 25, Fax: +98 21 66 02 32 61.}
  \fi
  }


%
\maketitle
\begin{abstract}
Let $\Ab$ be an $n \times m$ matrix with $m>n$, and suppose that the underdetermined linear system $\Ab \sbb = \xb$ admits a sparse solution $\sbb_0$ for which $\|\sbb_0\|_0 < \frac{1}{2} \spark(\Ab)$. Such a sparse solution is unique due to a well-known uniqueness theorem. Suppose now that we have somehow a solution $\hsbb$ as an estimation of $\sbb_0$, and suppose that $\hsbb$ is only `approximately sparse', that is, many of its components are very small and nearly zero, but not mathematically equal to zero. Is such a solution necessarily close to the true sparsest solution? More generally, is it possible to construct an upper bound on the estimation error $\|\hsbb-\sbb_0\|_2$ without knowing $\sbb_0$? The answer is positive, and in this paper we construct such a bound based on minimal singular values of submatrices of $\Ab$. We will also state a tight bound, which is more complicated, but besides being tight, enables us to study the case of random dictionaries and obtain probabilistic upper bounds. We will also study the noisy case, that is, where $\xb = \Ab \sbb+\nb$. Moreover, we will see that where $\nz{\sbb_0}$ grows, to obtain a predetermined guaranty on the maximum of $\nt{\hsbb-\sbb_0}$, $\hsbb$ is needed to be sparse with a better approximation. This can be seen as an explanation to the fact that the estimation quality of sparse recovery algorithms degrades where $\nz{\sbb_0}$ grows.
\end{abstract}


\IEEEpeerreviewmaketitle

\section{Introduction and problem statement}
\label{sec:intro}
\IEEEPARstart{S}{parse} solution of underdetermined systems of linear equations has recently attracted the attention of many researchers from different viewpoints, because of its potential applications in many different problems. It is used, for example, in Compressed Sensing (CS)~\cite{CandRT06,Dono06,Bara07}, underdetermined Sparse Component Analysis (SCA) and source separation~\cite{GribL06, BofiZ01, GeorTC04, LiCA03}, atomic decomposition on overcomplete dictionaries~\cite{ChenDS99,DonoET06},  decoding real field codes~\cite{CandT05}, image deconvolution~\cite{FiguN03,FiguN05}, image denoising~\cite{Elad06}, electromagnetic imaging and Direction of Arrival (DOA) finding~\cite{GoroR97}, etc. The importance of sparse solutions of underdetermined linear systems comes from the fact that although such systems have generally an infinite number of solutions, their sparse solutions may be {\em unique}. 

Let $\Ab=[\ab_1, \dots, \ab_m]$ be an $n \times m$ matrix with $m>n$, where $\ab_i$'s, $i=1,\dots,m$ denote its columns, and consider the Underdetermined System of Linear Equations (USLE)
\begin{equation}
	\Ab \sbb = \xb.
	\label{eq: noiseless system}
\end{equation}
By the sparsest solution of the above system one means a solution $\sbb$ which has as small as possible number of nonzero components. In signal (or atomic) decomposition viewpoint, $\xb$ is a signal which is to be decomposed as a linear combination of the signals $\ab_i$'s, $i=1,\dots,m$, and hence, $\ab_i$'s are usually called~\cite{MallZ93} `atoms', and $\Ab$ is called the `dictionary' over which the signal is to be decomposed. When the dictionary is overcomplete ($m>n$), the representation is not unique, but by the sparsest solution, we are looking for the representation which uses as small as possible number of atoms to represent the signal.

It has been shown~\cite{GoroR97,GribN03,DonoE03} that if (\ref{eq: noiseless system}) has a sparse enough solution, it is its {\em unique} sparsest solution. More precisely:
\begin{theorem}[Uniqueness Theorem~\cite{GribN03,DonoE03}]
Let $\spark(\Ab)$ denote the minimum number of columns of $\Ab$ that are linearly dependent, and $\|\cdot\|_0$ denotes the $\ell^0$ norm of a vector (i.e. the number of its nonzero components). Then if the USLE $\Ab \sbb = \xb$ has a solution $\sbb_0$ for which $\|\sbb_0\|_0 < \frac{1}{2} \spark(\Ab)$, it is its unique sparsest solution.
\end{theorem}
A special case of this uniqueness theorem has also been stated in~\cite{GoroR97}: if $\Ab$ satisfies the Unique Representation Property (URP), that is, if all $n \times n$ submatrices of $\Ab$ are non-singular, then $\spark(\Ab)=n+1$ and hence $\|\sbb_0\|_0 \le \frac{n}{2}$ implies that $\sbb_0$ is the unique sparsest solution.

Although the sparsest solution of (\ref{eq: noiseless system}) may be unique, finding this solution requires a combinatorial search and is generally NP-hard. Then, many different sparse recovery algorithms have been proposed to find an estimation of $\sbb_0$, for example, Basis Pursuit (BP)~\cite{ChenDS99}, Matching Pursuit (MP)~\cite{MallZ93}, FOCUSS~\cite{GoroR97}, Smoothed L0 (SL0)~\cite{MohiBJ07,MohiBJ09}, SPGL1~\cite{BergF08}, IDE~\cite{AminBJ06}, ISD~\cite{WangY09}, etc.

Now, consider the following two different cases:
\begin{itemize}
	\item Exact sparsity: We say that a vector $\sbb$ is sparse in the {\em exact sense} if many of its components are {\em exactly equal to zero}. More precisely, $\sbb$ is said to be $k$-sparse in the {\em exact sense} if it has at most $k$ nonzero entries (and all other entries are {\em exactly equal to zero}).
	
	\item Approximate Sparsity: We say that a vector $\sbb$ is sparse in the {\em approximate sense} if many of its components are very small and {\em approximately equal to zero} (but not necessarily `exactly' equal to zero). More precisely, $\sbb$ is said to be $k$-sparse with approximation $\epsilon$ if it has at most $k$ entries with magnitudes larger than $\epsilon$ (all of its other entries have magnitudes smaller than $\epsilon$).
\end{itemize}

Some of the sparse recovery algorithms (\eg\ BP based on Simplex linear programming) return estimations which are sparse in the exact sense, while some others (\eg\ MP with large enough iterations, SL0, FOCUSS and SPGL1) return solutions which are sparse only in the approximate sense.

Suppose now that by using any algorithm (or simply by a magic guess) we have found a solution $\hsbb$ of $\Ab \sbb = \xb$, as an estimation of the true sparsest solution ($\sbb_0$). The question now is: ``{\em Noting that $\sbb_0$ is unknown, is it possible to construct an upper bound for the estimation error $\|\hsbb-\sbb_0\|_2$ only from $\hsbb$, where $\nt{\cdot}$ stands for the $\ell^2$ norm}''? For example, if $\Ab$ satisfies the URP, and $\nz{\hsbb}$ is less than or equal $\lfloor n/2 \rfloor$, where $\lfloor x \rfloor$ stands for the largest integer smaller than or equal to $x$, then the uniqueness theorem insures that $\hsbb=\sbb_0$. On the other hand, if all the components of $\hsbb$ are nonzero but its $(\lfloor n/2 \rfloor+1)$'th largest magnitude component is very small, heuristically we expect to be close to the true solution $\sbb_0$, but the uniqueness theorem says nothing about this heuristic. 

In this paper, we will see that the answer to the above question is positive, and we will construct  
upper bounds on $\|\hsbb-\sbb_0\|_2$ without knowing $\sbb_0$, which depend on the matrix $\Ab$ and (in the case $\Ab$ satisfies the URP) are proportional to the magnitude of the $(\lfloor n/2 \rfloor+1)$'th largest component of $\hsbb$. Consequently, if the $(\lfloor n/2 \rfloor+1)$'th largest component of $\hsbb$ is zero, then our upper bounds vanish, and hence $\hsbb=\sbb_0$.
This is, in fact, the same result provided by the uniqueness theorem, and hence our upper bounds can be seen as a generalization of the uniqueness theorem. In other words,
from the classical uniqueness theorem, all that we know is that if among $m$ components of $\hsbb$, $m-\lfloor n/2 \rfloor$ components are `{\em exactly\/}' zero, then $\hsbb=\sbb_0$, but {\em if $\hsbb$ has more than $\lfloor n/2 \rfloor$ nonzero components (even if $m-\lfloor n/2 \rfloor$ of its components have very very small magnitudes) we are not sure to be close to the true solution}. As we will see in this paper, our upper bounds, however, insure that in the second case, too, we are not far from the true solution. Moreover, the dependence of our upper bounds on $\Ab$ provides some explanations about the sensitivity of the error to the properties of the matrix $\Ab$.

Constructing an upper bound on the error $\|\hsbb-\sbb_0\|_2$ can also be found in some other works, \eg~\cite{Ward09,MaliSW10,FoucL09}. In some of these works (\eg~\cite{Ward09,MaliSW10}) the bounds are probabilistic, that is, they have been obtained for random dictionaries and shown to be held with probabilities larger than certain values. Being non-deterministic, these bounds cannot be used to infer deterministic results. For example, they cannot be used to say whether or not the heuristic stated above (that is, ``if $\hsbb$ has at most $n/2$ `large' components, then it is close to the true solution'') is generally true or not, while our bounds answer this question. Another difference between our bounds with those of~\cite{Ward09,MaliSW10} is that in~\cite{Ward09,MaliSW10} it has been assumed that we have at hand an {\em algorithm} for estimating the sparsest solution of an underdetermined linear system and several calls to this algorithm are required, whereas in this paper, we have at hand only a {\em single estimation} ($\hsbb$) of the sparsest solution ($\sbb_0$), and we are going to develop upper bounds on the error $\nt{\hsbb-\sbb_0}$ without knowing $\sbb_0$. 
Moreover, the bounds in some of these works (\eg~\cite{MaliSW10,FoucL09}) have been constructed for specific methods used for finding the estimation $\hsbb$, \eg\ minimizing $\ell^1$ or $\ell^q$ norms for $0 < q \le 1$, whereas in this paper {\em we are discussing the bounds based on $\hsbb$ itself and independent of the method used for its estimation: it may be obtained by any algorithm or by a magic guess}. 
In fact, to our best knowledge, constructing a deterministic bound on $\|\hsbb-\sbb_0\|_2$ and independent of the method used for obtaining $\hsbb$ has not previously been addressed in the literature. 
Note however that although our deterministic bounds can be used to infer deterministic results, they are not suitable for practical calculation, because they need Asymmetric Restricted Isometry Constant (ARIC)~\cite{FoucL09,DaviG09} of a dictionary, or similar quantities, whose calculation are computationally intractable for large matrices (note however that these quantities have to be calculated only once for each dictionary). We will also present a probabilistic bound for random dictionaries, which is again independent of the method used to obtain the estimate $\hsbb$.

A related problem has already been addressed in~\cite{GribVV06}, in which, for the noisy case $\xb = \Ab \sbb + \eb$, deterministic upper bounds have been constructed for the error $\| \hsbb - \sbb_0\|_q$ (for a set of different $q$'s including $q=2$). However, in that paper it has been implicitly assumed that $\hsbb$ is sparse in the exact sense, that is, $\|\hsbb\|_0 \le \lfloor n/2 \rfloor$, otherwise, their upper bounds grow to infinity. On the other hand, if the noise power ($\|\eb\|^2$) is set equal to zero, the upper bounds of~\cite{GribVV06} for $\| \hsbb - \sbb_0\|$ vanish, resulting again to the uniqueness theorem. In other words, reference~\cite{GribVV06} can be seen somehow as a generalization of the uniqueness theorem to the noisy case, whereas our paper can be seen as a generalization of the uniqueness theorem to the case $\hsbb$ is not sparse in the exact sense. We will also consider in Section~\ref{sec: noisy} the case where there is noise {\em and} $\hsbb$ is sparse in the approximate sense. Some error bounds for the noisy case have also been obtained in~\cite{DonoET06}, but those bounds are for specific algorithms for estimating $\sbb_0$, while our bounds are only based on $\hsbb$ itself and independent of the method used for finding it.

Some parts of this work have been presented in the conference paper~\cite{BabaMJ09}. 
Here, we study the problem more thoroughly (without repeating some details of that conference paper), and we provide also a tight bound on the above error. Imposing no assumption on the normalization of the columns of the dictionary, this tight bound will enable us to obtain a probabilistic upper bound.
Moreover, we address the noisy case where $\hsbb$ is sparse in the approximate sense.

The paper is organized as follows. In Section~\ref{sec: main idea} we review a first result already stated in~\cite{MohiBJ09}, which provides the basic idea of this paper. Then in Section~\ref{sec: sigmin bound}, we present a bound based on minimal singular values of the submatrices of the dictionary. Our tight bound is then presented in Section~\ref{sec: tight bound}. By considering the noisy case in Section~\ref{sec: noisy}, we complete our discussion on deterministic dictionaries before studying random dictionaries in Section~\ref{sec: random}.

\section{A first bound}\label{sec: main idea}
A first result has been given in Corollary~1 of Lemma~1 of \cite{MohiBJ09} during the analysis of the convergence of the SL0 algorithm. We review that result here (with a few changes in notations).

For the $n \times m$ matrix $\Ab$, let ${\cal P}_j(\Ab)$, $1\le j \le m$, denote the set of all matrices which are obtained by taking $j$ columns of $\Ab$. Moreover, let ${\cal M}_n(\Ab)\triangleq{\cal P}_1(\Ab) \cup {\cal P}_2(\Ab) \cup \cdots \cup {\cal P}_n(\Ab)$, and define
\begin{equation}
	\ga \triangleq \max_{\Bb \in {\cal M}_n(\Ab)} \|\Bpinv\|_F,
\end{equation}
where $\Bpinv$ stands for the Moore-Penrose pseudoinverse of $\Bb$, and $\| \cdot \|_F$ denotes the Frobenius norm of a matrix. The constant $\ga$ depends only on the dictionary $\Ab$. Moreover, for a vector $\yb$ and a positive scalar $\alpha$, let $\|\yb\|_{0, \alpha}$ denote the number of components of $\yb$ which have magnitudes larger than $\alpha$. In other words, $\|\yb\|_{0, \alpha}$ denotes the $\ell^0$ norm of a thresholded version of $\yb$ in which the components with magnitudes smaller than or equal to $\alpha$ are clipped to zero.

The Corollary~1 of Lemma~1 of~\cite{MohiBJ09} states then:

\begin{coro}[of~\cite{MohiBJ09}]\label{coro: Mohimani}
Let $\Ab$ be an ${n \times m}$ matrix with unit $\ell^2$ norm columns which satisfies the URP and let $\deltab\in\mathrm{null}(\Ab)$. If for an $\alpha>0$, $\deltab$ has at most $n$ components with absolute values greater than $\alpha$ (that is, if $\|\deltab\|_{0, \alpha} \le n$), then
\begin{equation}\label{eq: first lem}
	\|\deltab\|_2<(\ga+1)m\alpha.
\end{equation}
\end{coro}

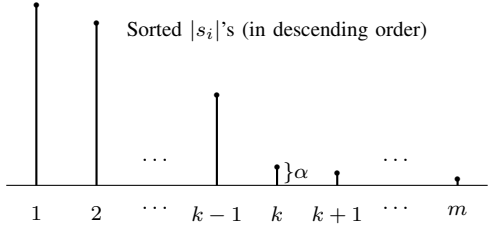
\begin{figure}
  \unitlength 0.8cm
  \centering
  \begin{picture}(9.5, 4.4)(0, -0.9)
  \thinlines
  \put(0.5,0){\line(1,0){8}}

  \thicklines

  \put(1, 0){\line(0, 1){3}}
  \put(1,3){\circle*{0.1}}
  \put(0.5, -1.35){\makebox(1, 1)[t]{\footnotesize $1$}}

  \put(2, 0){\line(0, 1){2.7}}
  \put(2,2.7){\circle*{0.1}}
  \put(1.5, -1.35){\makebox(1, 1)[t]{\footnotesize $2$}}

  \put(2.5, -1.35){\makebox(1, 1)[t]{\footnotesize $\dots$}}
  \put(2.5, 0.4){\makebox(1, 1)[b]{\footnotesize $\dots$}}

  \put(4, 0){\line(0, 1){1.5}}
  \put(4,1.5){\circle*{0.1}}
  \put(3.5, -1.35){\makebox(1, 1)[t]{\footnotesize $k-1$}}
  \put(4.5, 2.4){\makebox(1, 1)[b]{\footnotesize Sorted $|s_i|$'s (in descending order)}}

  \put(5, 0){\line(0, 1){0.3}}
  \put(5,0.3){\circle*{0.1}}
  \put(4.5, -1.35){\makebox(1, 1)[t]{\footnotesize $k$}}
  \put(5.1, 0.03){\makebox(1, 0.3)[bl]{\footnotesize $\}\alpha$}}

  \put(6, 0){\line(0, 1){0.2}}
  \put(6,0.2){\circle*{0.1}}
  \put(5.5, -1.35){\makebox(1, 1)[t]{\footnotesize $k+1$}}

  \put(6.5, -1.35){\makebox(1, 1)[t]{\footnotesize $\dots$}}
  \put(6.5, 0.4){\makebox(1, 1)[b]{\footnotesize $\dots$}}

  \put(8, 0){\line(0, 1){0.1}}
  \put(8,0.1){\circle*{0.1}}
  \put(7.5, -1.35){\makebox(1, 1)[t]{\footnotesize $m$}}

\end{picture}

\caption{The definition of $h(k,\sbb)$: Sort the magnitudes of the entries of $\sbb$ in descending order. 
Then, $h(k,\sbb)$ is the magnitude of the $k$'s element (denoted by $\alpha$ in the figure).}
\label{fig: alpha definition}

\end{figure}

\noindent We define now the following notation (see also Fig.~\ref{fig: alpha definition}):
\begin{defi}
Let $\sbb$ be a vector of length $m$. Then $h(k,\sbb)$ denotes the magnitude of the $k$'th largest magnitude component of $\sbb$.
\end{defi}

Then, using the above corollary, Remark~5 of Theorem~1 of~\cite{MohiBJ09} states the following idea to construct an upper bound on $\|\hsbb - \sbb_0\|_2$ as follows: Let 
$\alpha_{\hsbb,n} \triangleq h(\lfloor \frac{n}{2}\rfloor+1, \hsbb)$. Since the true sparsest solution ($\sbb_0$) has at most $\lfloor \frac{n}{2} \rfloor$ nonzero components, $\hsbb-\sbb_0$ has at most $n$ 
components with absolute values greater than $\alpha_{\hsbb,n}$, that is, $\|\hsbb-\sbb_0\|_{0, \alpha_{\hsbb,n}} \le n$. 
Moreover, $(\hsbb-\sbb_0)\in\mathrm{null}(\Ab)$ and 
hence Corollary~\ref{coro: Mohimani} implies that
\begin{equation}
  \label{eq: Mohi bound}
	\|\hsbb-\sbb_0\|_2 \le (\ga+1)m \alpha_{\hsbb,n}\, .
\end{equation}
This result is consistent with the heuristic stated in the introduction: ``{\em if $\hsbb$ has at most
$n/2$ `large' components, the uniqueness of the sparsest solution insures that $\hsbb$
is close to the true solution}''.

\section{A bound based on minimal singular values}\label{sec: sigmin bound}
The bound (\ref{eq: Mohi bound}) is not easy to be analyzed and worked with. Especially, the dependence of the bound on the dictionary (through the constant $\ga$) is very complicated. Moreover, calculating the $\ga$ constant for a dictionary requires calculation of the pseudoinverses of all of the $\binom{m}{1} + \binom{m}{2} + \cdots + \binom{m}{n}$ elements of ${\cal M}_n(\Ab)$. In this section, we modify (\ref{eq: Mohi bound}) to obtain a bound that is easier to be analyzed and (in a statistical point of view) its dependence to (the statistics of) $\Ab$ is simpler. Moreover, we state our results for more general cases than where $\Ab$ satisfies the URP.

\subsection{Definitions and notations}
For a matrix $\Bb$ let $\sigma_{\min}(\Bb)$ or $\sigma_{\min, \Bb}$ denote its smallest singular value%
\footnote{In some references, \eg~\cite{Hayk96}, the singular values of a matrix are defined to be {\em strictly positive} quantities. This definition is not appropriate for this paper. We are using the more common definition of Horn and Johnson~\cite[pp. 414-415]{HornJ90}, in which, the singular values of a $p \times q$ matrix $\Mb$ are the square roots of the $\min(p,q)$ largest eigenvalues of $\Mb^H \Mb$ (or $\Mb \Mb^H$). Using this definition, there are always $\min(p,q)$ singular values, where a zero singular value characterizes a (tall or wide) non-full-rank matrix.}%
. Similarly, we denote its largest singular value by $\sigma_{\max}(\Bb)$ or $\sigma_{\max, \Bb}$.
We now define the following notations about the dictionary $\Ab$:

\begin{itemize}
	\item Let $q=q(\Ab)=\spark(\Ab)-1$. Then, by definition, any $q$ columns of $\Ab$ are linearly independent, and there is at least one set of $q+1$ columns which are linearly dependent (in the literature, the quantity $q$ is usually called `Kruskal rank' or `k-rank' of $\Ab$). It is also obvious that $q\le n$, in which, $q=n$ corresponds to the case where $\Ab$ satisfies the URP.

	\item Let $\sm{j}{\Ab}$ or $\varsm{j}{\Ab}$ denote the smallest singular value among the singular values of all submatrices of $\Ab$ obtained by taking $j$ columns of $\Ab$, that is,
\begin{equation} \label{eq: sm def}
	 \sm{j}{\Ab} = \min_{\Bb \in {\cal P}_j(\Ab)} \{\sigma_{\min}(\Bb)\}\cdot
\end{equation}
\end{itemize}

Note that since any $q$ columns of $\Ab$ are linearly independent, we have $\sm{j}{\Ab} > 0$, for all $1\le j \le q(\Ab)$.

Recall now the following lemma~\cite[p. 419]{HornJ90} (we presented a direct simple proof for the first two parts of this lemma in~\cite{BabaMJ09}).
\begin{lemma}\label{lem: sing val B b}
Let $\Bb$ be an $n \times p$ matrix, and let $\Bb'$ denote the matrix obtained by adding a new column to $\Bb$. Then:
\begin{itemize}

\item[a)]
 If $p < n$ ($\Bb$ is tall), then $\sigma_{\min}(\Bb')\le \sigma_{\min}(\Bb)$.

\item[b)]
 If $p \ge n$ ($\Bb$ square or wide), then $\sigma_{\min}(\Bb')\ge \sigma_{\min}(\Bb)$.
 
\item[c)] We have always $\sigma_{\max}(\Bb')\ge \sigma_{\max}(\Bb)$.
\end{itemize}
\end{lemma}

Using the above lemma, the sequence $\varsm{j}{\Ab}$, $j=1,\dots,m$ is decreasing for $1 \le j \le q$ and increasing for $n \le j \le m$. More precisely, if $q=n$ (URP case), we have
\begin{equation}
	\varsm{1}{\Ab} \ge \varsm{2}{\Ab} \ge \cdots \ge \underbrace{\varsm{n}{\Ab}}_{>0} \le \varsm{n+1}{\Ab} \le \cdots \le \varsm{m}{\Ab},
	\label{eq: sig decrease q eq n}
\end{equation}
and if $q<n$, we have
\ifonecol
\begin{equation}
	\varsm{1}{\Ab} \ge \cdots \ge \varsm{q}{\Ab} >0 = \varsm{q+1}{\Ab} = \cdots = \varsm{n}{\Ab} \le \varsm{n+1}{\Ab} \le \cdots \le \varsm{m}{\Ab} \cdot
	\label{eq: sig decrease q lt n}
\end{equation}
\else
\begin{equation}
 \begin{split}
	\varsm{1}{\Ab} \ge \cdots & \ge \varsm{q}{\Ab} >0 = \varsm{q+1}{\Ab} = \\
	&= \cdots = \varsm{n}{\Ab} \le \varsm{n+1}{\Ab} \le \cdots \le \varsm{m}{\Ab} \cdot
 \end{split}
 \label{eq: sig decrease q lt n}
\end{equation}
\fi

\noindent Note also that (\ref{eq: sig decrease q eq n}) and (\ref{eq: sig decrease q lt n}) imply that for $1 \le j \le q$
\begin{equation}
\sm{j}{\Ab}= \min_{\|\xb\|_0 \le j} {\|\Ab \xb\|_2}/{\|\xb\|_2} \cdot
\label{eq: ARIC form}
\end{equation}

\noindent {\bf Remark. }
The quantity defined in (\ref{eq: sm def}) is closely related to Restricted Isometry Property (RIP)~\cite{CandT05,Cand08}, and is in fact the left Asymmetric Restricted Isometric Constant (ARIC) of $\Ab$~\cite{FoucL09,DaviG09}. As introduced in~\cite{CandT05}, 
the Restricted Isometry Constant (RIC) of $\Ab$ is defined as the smallest $\delta_j$ such that $(1-\delta_j)\nt{\xb}^2 \le \nt{\Ab \xb}^2\le (1+\delta_j)\nt{\xb}^2$ for all vectors $\xb\in\Rr^m$ with $\nz{\xb}\le j$. The lower and upper bounds of this inequality are symmetric, and hence the authors of~\cite{FoucL09,DaviG09} introduced asymmetric RIC's, which are defined as the best $\alpha_j$ and $\beta_j$ such that $\alpha_j\nt{\xb} \le \nt{\Ab \xb}\le \beta_j\nt{\xb}$ for all vectors $\xb\in\Rr^m$ with $\nz{\xb}\le j$. Comparing with (\ref{eq: ARIC form}), it is seen that the left ARIC ($\alpha_j$) is the same quantity denoted by $\sm{j}{\Ab}$ in above.

\subsection{The upper bound}
Now we state the main theorem of this section:
\begin{theorem}\label{th: sigmin bound}
Let $\Ab$ be an ${n \times m}$ matrix ($m>n$)  with unit $\ell^2$ norm columns.
Suppose that $\sbb_0$ is a solution of $\Ab \sbb = \xb$ for which $\|\sbb_0\|_0 \le {\ell}/{2}$, where $\ell$ is an arbitrary integer less than or equal to $q(\Ab)$.
Let $\hsbb$ be a solution of $\Ab \sbb = \xb$, and define $\alpha_{\hsbb,\ell}\triangleq h(\lfloor {\ell}/{2} \rfloor+1,\hsbb)$. Then
\begin{equation}
  \label{eq: sigmin bound}
	\|\hsbb-\sbb_0\|_2 \le \left(\frac{1}{\varsm{\ell}{\Ab}}+1 \right) m \, \alpha_{\hsbb,\ell}\, .
\end{equation}
\end{theorem}

\medskip

Before going to the proof, let us state a few remarks on the consequences of the above theorem.

\medskip

\noindent {\bf Remark 1. } 
Suppose that $\Ab \sbb = \xb$ has a sparse solution $\sbb_0$ which satisfies $\nz{\sbb_0} \le \frac{1}{2} q(\Ab)$. By setting $\ell=q(\Ab)$ in (\ref{eq: sigmin bound}), which is the largest $\ell$ satisfying the conditions of the theorem, we will have
\begin{equation}
  \label{eq: sigmin bound spark}
	\|\hsbb-\sbb_0\|_2 \le \left(\frac{1}{\varsm{q}{\Ab}}+1 \right) m \, \alpha_{\hsbb,q}\, .
\end{equation}
If the estimated sparse solution $\hsbb$ satisfies also $\nz{\hsbb} \le \frac{q}{2}$, then $\alpha_{\hsbb,q}=0$, hence the upper bound in (\ref{eq: sigmin bound}) vanishes, and therefore $\hsbb=\sbb_0$. In other words, the above theorem implies that a solution with $\|\sbb\|_0 \le \frac{1}{2}q(\Ab)$ is unique, that is, the above theorem implies the uniqueness theorem. For example, for the special case of $\Ab$ satisfying the URP ($q(\Ab)=n$), if we have found a solution satisfying $\|\hsbb\|_0 \le \frac{n}{2}$, we are sure that we have found the unique sparsest solution.

\medskip

\noindent {\bf Remark 2. } 
Moreover, if the estimated sparse solution $\hsbb$ is sparse only in the approximate sense, that is, if $m-\lfloor \frac{q}{2} \rfloor$ components of $\hsbb$ have very small magnitudes, then $\alpha_{\hsbb,q}$ is small, and the bound (\ref{eq: sigmin bound spark}) states that we are probably (depending on the matrix $\Ab$) close to the true solution. Moreover, in this case, $\varsm{q}{\Ab}$ determines some kind of sensitivity to the dictionary: For example, if the URP holds ($q=n$) but there exists an $n\times n$ square submatrix of $\Ab$ which is ill-conditioned, then $\varsm{n}{\Ab}$ is very small and hence for achieving a predetermined accuracy, $\alpha_{\hsbb,n}$ should be very small, that is, the sparsity of $\hsbb$ should be held with a better approximation.

\medskip 

\noindent {\bf Remark 3. } 
Theorem~\ref{th: sigmin bound} states also some kind of `sensitivity' to the degree of sparseness of the sparsest solution $\sbb_0$. Let $p \triangleq \|\sbb_0\|_0$, and set $\ell=2p$ in (\ref{eq: sigmin bound}), and suppose that $\ell \le q(\Ab)$. Then the conditions of Theorem~\ref{th: sigmin bound} have been satisfied and hence (\ref{eq: sigmin bound}) becomes
\begin{equation}
  \|\hsbb-\sbb_0\|_2 \le \left(\frac{1}{\varsm{2p}{\Ab}}+1 \right) m \, \alpha_{\hsbb,2p} \, .
  \label{eq: sensitivity}
\end{equation}
In other words, whenever $\sbb_0$ is sparser, $p$ is smaller, hence from (\ref{eq: sig decrease q eq n}) and (\ref{eq: sig decrease q lt n}) $\varsm{2p}{\Ab}$ is larger, and therefore a larger $\alpha_{\hsbb,2p}$ is tolerable (that is, we have less sensitivity to exact sparseness of $\hsbb$). This can somehow explain the fact that sparse recovery algorithms work better for sparser ${\sbb_0}$'s~\cite{MohiBJ09}.

\subsection{Proof}
To prove Theorem~\ref{th: sigmin bound} we first state a modified version of (\ref{eq: first lem}):

\begin{prop}\label{prop: sigmin bound}
Let $\Ab$ be an ${n \times m}$ matrix ($m>n$)  with unit $\ell^2$ norm columns, and assume
that any $\ell$ columns of $\Ab$ are linearly independent ($\ell \le n$). Let $\deltab \in \mathrm{null}(\Ab)$. If for an $\alpha\ge 0$, $\|\deltab\|_{0,\alpha} \le \ell$, then
\begin{equation}\label{eq: prop sigmin}
\|\deltab\|_2 \le \left( \frac{1}{\varsm{\ell}{\Ab}} + 1 \right) m \alpha.
\end{equation}
\end{prop}

\begin{IEEEproof}
The proof is based on some modifications to the proof of Lemma~1 of~\cite{MohiBJ09}.

Let $m_l$ be the number of components of $\deltab$ with magnitudes larger than $\alpha$, and $m_s$ be the number of components of $\deltab$ with magnitudes smaller than or equal to $\alpha$. In other words, $m_l=\|\deltab\|_{0,\alpha}$ and $m_s=m-\|\deltab\|_{0,\alpha}$ (indexes $l$ and $s$ in $m_l$ and $m_s$ stand for `\underline{L}arge' and `\underline{S}mall'). We consider two cases:

{\bf Case 1} ($m_l=0$ and $m_s=m$): In this case, all the components of $\deltab$ have magnitudes less than or equal to $\alpha$, and hence we can simply write $\|\deltab\|_2 \le \sum_i |\delta_i| \le m\alpha$ which satisfies also (\ref{eq: prop sigmin}).

\medskip 

{\bf Case 2} ($1 \le m_l\le \ell$ and  $m_s\le m-1$): In this case, there is at least one component of $\deltab$ with magnitude larger than $\alpha$. Let $\deltab_l$ be composed of the components of $\deltab$ which have magnitudes larger than $\alpha$, and $\Ab_l$ be composed of the corresponding columns of $\Ab$. Similarly\footnote{\label{footnote: Al and As}In MATLAB notation $\Ab_l\triangleq\Ab(:, {\tt abs}(\deltab)\!>\!\alpha)$, $\Ab_s\triangleq\Ab(:, {\tt abs}(\deltab)\!\le\!\alpha)$.}, let $\deltab_s$ be composed of the components of $\deltab$ which have magnitudes less than or equal to $\alpha$, and $\Ab_s$ is composed of the corresponding columns of $\Ab$.
Since $\deltab \in \mathrm{null}(\Ab)$, ${\bf 0}= \Ab \deltab = \Ab_l \deltab_l + \Ab_s \deltab_s$, and define
\begin{equation}
	\bb \triangleq \Ab_l \deltab_l = - \Ab_s \deltab_s.
\end{equation}
From $\bb = - \Ab_s \deltab_s$,
\begin{gather}
  \|\bb\|_2 = \|\Ab_s \deltab_s\|_2 = \|\sum_i \delta_{s,i} \ab_{s,i} \|_2 \le \sum_i \underbrace{|\delta_{s,i}|}_{\leq \alpha}\underbrace{\|\ab_{s,i}\|_2}_{1} \nonumber \\
  \Rightarrow \|\bb\|_2 \le m_s \alpha \le  (m-1) \alpha \le m \alpha. \label{eq: prop nontight up b}
\end{gather}

\noindent From $\bb = \Ab_l \deltab_l$,
\begin{gather}
	\|\bb\|_2 = \|\Ab_l \deltab_l\|_2 \ge \sigma_{\min}(\Ab_l) \|\deltab_l\|_2 \nonumber \\
	\Rightarrow \|\deltab_l\|_2 \le \frac{\|\bb\|_2}{\sigma_{\min}(\Ab_l)}\cdot \label{eq: prop nontight up d_l based on b}
\end{gather}
Note that in the above equations, the assumption $m_l\le \ell$ was essential, otherwise $\|\Ab_l \deltab_l\|_2$ and $\sigma_{\min}(\Ab_l)$ could be zero. Combining now (\ref{eq: prop nontight up b}) and (\ref{eq: prop nontight up d_l based on b}), we will have
\begin{equation}
	\|\deltab_l\|_2 \le \frac{m\alpha}{\sigma_{\min}(\Ab_l)}\cdot \label{eq: prop nontight up d_l}
\end{equation}
Moreover, $\|\deltab_s\|_2\le m_s \alpha \le (m-1)\alpha \le m \alpha$. Therefore
\begin{equation}
	\|\deltab\|_2 \le \|\deltab_l\|_2 + \|\deltab_s\|_2 \le \frac{m\alpha}{\sigma_{\min}(\Ab_l)} + m \alpha.
	\label{eq: prop nontight up delta}
\end{equation}
Now, from the definition (\ref{eq: sm def}) and Lemma~\ref{lem: sing val B b}, $\sigma_{\min}(\Ab_l) \ge \sm{m_l}{\Ab_l} \ge \sm{\ell}{\Ab_l}$, which proves the proposition.
\end{IEEEproof}

\medskip 

\begin{IEEEproof}[Proof of Theorem~\ref{th: sigmin bound}]
 $\sbb_0$ has at most $\lfloor \frac{\ell}{2} \rfloor$ nonzero components and $\hsbb$ has at most $\lfloor \frac{\ell}{2} \rfloor$ components with magnitudes larger than $\alpha$. Therefore, $\hsbb-\sbb_0$ has at most $\ell$ components with magnitudes larger than $\alpha$. Moreover, $(\hsbb-\sbb_0)\in\mathrm{null}(\Ab)$. Hence, the conditions of Proposition~\ref{prop: sigmin bound} hold for $\deltab=\hsbb-\sbb_0$ and $\alpha=\alpha_{\hsbb,\ell}$, which proves the theorem.
\end{IEEEproof}

\section{A tight bound}\label{sec: tight bound}
Although the bound in (\ref{eq: sigmin bound}) is relatively simple, except for the trivial case $\alpha_{\hsbb,\ell}=0$ the equality in (\ref{eq: sigmin bound}) can never be satisfied (as will be explained in the proof of Theorem~\ref{th: gamma bound}). Therefore, for an approximate sparse $\hsbb$ the bound in (\ref{eq: sigmin bound}) is not tight. In this section we present a tight bound on the estimation error $\hsbb - \sbb_0$, which does not depend only on minimal singular values of submatrices of $\Ab$, but depends on a quantity $\bar\gamma(\Ab)$ defined below.

\subsection{Definitions and notations}\label{sec: eta defi}
\begin{defi}
Let $\Ab$ be an $n \times m$ matrix and $m>n$. For any $\Bb\in \Pc_j(\Ab)$, let $\Bbbar$ denote the matrix composed of the columns of $\Ab$ which are not in $\Bb$. For $j=1,\dots,q(\Ab)$ we define
\begin{equation}
\eta_j(\Ab)\triangleq \max_{\Bb \in \Pc_j(\Ab)} \left\{	\frac{\sigma_{\max}(\Bbbar)}{\sigma_{\min}(\Bb)} \right\}\cdot
\label{eq: eta def}
\end{equation}
\end{defi}

Note that while $j\le q(\Ab)$, $\sigma_{\min}(\Bb)>0$, and hence $\eta_j(\Ab)<\infty$ for $j=1,\dots,q(\Ab)$.

\begin{defi}\label{defi: gamma}
Let matrix $\Ab$ be as in the previous definition. For $j=1,\dots,q(\Ab)$ we define the quantities
\begin{gather}
\gamma_j(\Ab) \triangleq  \sqrt{(m-j) \, \Big(1+\eta^2_j(\Ab)\Big)}, \label{eq: gamma def} \\
\bar\gamma_j(\Ab)\triangleq \max \Big\{ \gamma_1(\Ab), \gamma_2(\Ab), \dots, \gamma_j(\Ab) \Big\}, \\
\bar\gamma'_j(\Ab)\triangleq \max \Big\{ \sqrt{m},\gamma_1(\Ab), \gamma_2(\Ab), \dots, \gamma_j(\Ab) \Big\}\cdot
\end{gather}
We also use the notations $\bar\gamma(\Ab)$ and $\bar\gamma'(\Ab)$ to denote the largest $\gamma_j(\Ab)$ and $\gamma_j'(\Ab)$ over the whole range of $j$, that is, $\bar\gamma(\Ab) \triangleq \bar\gamma_q(\Ab)=\max\{\gamma_1(\Ab), \dots, \gamma_q(\Ab)\}$, and $\gamma'(\Ab)=\max\{\sqrt{m},\gamma(\Ab)\}$.
\end{defi}

\medskip

\noindent {\bf Remark. } 
Note that the sequence $\eta_j(\Ab)$, $j=1,\dots,n$ is not necessarily increasing. In effect, by taking one column from a matrix $\Mb$ and appending it to a matrix $\Nb$, the ratio ${\sigma^2_{\max}(\Mb)}/{\sigma^2_{\min}(\Nb)}$ does not necessarily increase, because both of its numerator and denominator decrease by Lemma~\ref{lem: sing val B b}. As an example, for the matrix \Ab=[ 0.79, 0.82, -0.84, -0.82; -0.57, 0.55, 0.33, -0.38; 0.23, -0.19, -0.43, 0.42], which has normalized columns (up to 2 decimal points), the sequence $\eta_j$ is approximately equal to $\{1.49,6.91,5.04\}$. Similarly, the sequence $\gamma_j(\Ab)$ is not necessarily increasing. For the above matrix, this sequence is approximately equal to $\{3.11,9.87,5.14\}$. However, as we will see in Section~\ref{sec: random}, for large random matrices with independently and identically distributed (iid) Gaussian entries, these sequences are both almost surely  increasing (see Remark~3 after Lemma~\ref{lem: g increasing}).

\subsection{The upper bound}

Now we are ready to state the main theorem of this section, which provides a tight upper bound on $\|\hsbb-\sbb_0\|_2$:

\begin{theorem}\label{th: gamma bound}
Let $\Ab$ be an ${n \times m}$ matrix ($m>n$), and suppose that $\sbb_0$ is a solution of $\Ab \sbb = \xb$ for which $\|\sbb_0\|_0 \le {\ell}/{2}$, where $\ell$ is an arbitrary integer less than or equal to $q(\Ab)$.
Let $\hsbb$ be a solution of $\Ab \sbb = \xb$, and define $\alpha_{\hsbb,\ell}\triangleq h(\lfloor {\ell}/{2} \rfloor+1,\hsbb)$. Then
\begin{equation}
	\|\hsbb-\sbb_0\|_2 \le \bar\gamma'_\ell(\Ab) \cdot \alpha_{\hsbb,\ell}\, .
	\label{eq: tight bound nonnorm}
\end{equation}
Moreover, if the columns of $\Ab$ are of unit $\ell^2$ norm, then
\begin{equation}
	\|\hsbb-\sbb_0\|_2 \le \bar\gamma_\ell(\Ab) \cdot \alpha_{\hsbb,\ell}\, .
	\label{eq: tight bound}
\end{equation}
\end{theorem}

\medskip

\noindent {\bf Remark 1. } 
For deterministic dictionaries, one usually uses normalized atoms, for which (\ref{eq: tight bound}) holds. However, (\ref{eq: tight bound nonnorm}) will be useful for random dictionaries (Section~\ref{sec: random}) with independently and identically distributed (iid) entries, for which having the unit $\ell^2$ norm cannot be guaranteed. Note also that if we normalize the columns of such a random dictionary, its entries would no longer be independent.

\smallskip

\noindent {\bf Remark 2. } 
In particular, by setting $\ell=q$ in (\ref{eq: tight bound}), the bound (\ref{eq: sigmin bound spark}) is modified to $\|\hsbb-\sbb_0\|_2 \le \bar\gamma(\Ab) \cdot \alpha_{\hsbb,q}$, and for the URP case
\begin{equation}
  \label{eq: tight bound URP}
	\|\hsbb-\sbb_0\|_2 \le \bar\gamma(\Ab) \cdot \alpha_{\hsbb,n}\, .
\end{equation}
In other words, if we know the constant $\bar\gamma(\Ab)$ for our dictionary, its multiplication by the $(\lfloor n/2 \rfloor+1)$'th largest magnitude component of the estimation $\hsbb$ would be an upper bound on the estimation error $\nt{\hsbb-\sbb_0}$. Like in (\ref{eq: sigmin bound}), this insures that if we have obtained an approximate sparse solution of $\Ab \sbb=\xb$, we are probably close to the true sparsest solution. However, unlike (\ref{eq: sigmin bound}) the bound in (\ref{eq: tight bound URP}) is tight. We will see in Section~\ref{sec: Tight Example} an example for which the equality in (\ref{eq: tight bound URP}) is satisfied.

\subsection{Proof}
From the proof of Theorem~\ref{th: sigmin bound}, it can be seen that if we obtain an upper bound for $\nt{\deltab}$ where $\deltab \in \mathrm{null}(\Ab)$ satisfies $\|\deltab\|_{0,\alpha} \le \ell$ for an $\alpha\ge 0$, we will obtain a bound for $\nt{\hsbb-\sbb_0}$ for $\alpha=\alpha_{\hsbb,\ell}$. Such a bound is given in the following proposition, as a modification to Proposition~\ref{prop: sigmin bound}.

\begin{prop}\label{prop: tight}
Let $\Ab$ and $\ell$ be as in Theorem~\ref{th: gamma bound}. Let $\deltab \in \mathrm{null}(\Ab)$, and suppose that for an $\alpha\ge 0$, $\|\deltab\|_{0,\alpha} \le \ell$. Let also $m_l$ and $m_s$ denote the number of entries of $\deltab$ with magnitudes larger than, and less than or equal to $\alpha$, respectively.

a) Case $m_l=0$ and $m_s=m$: (\ie\ where all entries of $\deltab$ have magnitudes smaller than or equal to $\alpha$). Then
\begin{equation}\label{eq: tight lemma a}
\|\deltab\|_2 \le \sqrt{m} \, \alpha.
\end{equation}

b) Case $1 \le m_l\le \ell$ and  $m_s\le m-1$: 
Let $\Ab_l$ and $\Ab_s$ be composed of the columns of $\Ab$ which correspond to the entries of $\deltab$ that have magnitudes larger than $\alpha$, and less than or equal $\alpha$, respectively (see footnote~\ref{footnote: Al and As}). Then
\begin{equation}\label{eq: tight prop}
\|\deltab\|_2 \le \left( 
\sqrt{m_s \left( 1 + \frac{\sigma^2_{\max}(\Ab_s)}{\sigma^2_{\min}(\Ab_l)}\right)}
 \, \right) \alpha.
\end{equation}
\end{prop}

\medskip

\begin{IEEEproof} The bound in (\ref{eq: prop sigmin}) is not tight because several inequalities used in its proof are not tight. Indeed, the equalities of the last inequality in (\ref{eq: prop nontight up b}) and also the first inequality in (\ref{eq: prop nontight up delta}) can never be met (unless for the trivial case $\alpha=0$). Hence, we prove the proposition by modifying the proof of Proposition~\ref{prop: sigmin bound}. Moreover, as opposed to what had been done in (\ref{eq: prop nontight up delta}), in this proof we do not use the assumption that the columns of $\Ab$ are normalized.

Note first that for a vector $\yb=(y_1, \dots, y_p)^T$, if $\forall i: 0 \le y_i \le \alpha$, then
\begin{equation}
	\nt{\yb} \le \sqrt{p} \, \alpha,
	\label{eq: unconst max norm y}
\end{equation}
and the equality holds if and only if $\forall i: y_i=\alpha$.

{\em Proof of Part a (case $m_l=0$ and $m_s=m$)}: In this case, (\ref{eq: unconst max norm y}) directly implies (\ref{eq: tight lemma a}), where the equality holds if and only if $|\delta_i|=\alpha\Rightarrow \delta_i=\pm \alpha$. Moreover, for the equality being satisfied, $\deltab$ has to be in $\mathrm{null}(\Ab)$, that is, $\Ab \deltab = \sum_i {\delta_i \ab_i} = \bf 0$, where $\ab_i$'s are the columns of $\Ab$. Hence, the upper bound in (\ref{eq: tight lemma a}) is tight, and is achieved only for the dictionaries for which a linear combination of their columns with $+1$ or $-1$ coefficients vanishes ($\sum {\pm \ab_i} = \bf 0$).

{\em Proof of Part b (case $m_l\ge 1, m_s\le m-1$)}: We follow the same argument as in the proof of Proposition~\ref{prop: sigmin bound}, but instead of (\ref{eq: prop nontight up b}) we write
\begin{equation}
  \|\bb\|_2 = \|\Ab_s \deltab_s\|_2 \le \sigma_{\max}(\Ab_s) \nt{\deltab_s}\cdot
  \label{eq: prop tight up b}
\end{equation}
Combining it with (\ref{eq: prop nontight up d_l based on b}), we will have (instead of (\ref{eq: prop nontight up d_l}))
\begin{equation}
	\|\deltab_l\|_2 \le \frac{\sigma_{\max}(\Ab_s)}{\sigma_{\min}(\Ab_l)} \nt{\deltab_s} \label{eq: prop tight up d_l}\cdot
\end{equation}
Finally, instead of (\ref{eq: prop nontight up delta}) we write $\|\deltab\|^2_2 = \|\deltab_s\|^2_2 + \|\deltab_l\|^2_2$, and hence from the above inequality we will have
\begin{equation}
	\nt{\deltab}^2 \le \nt{\deltab_s}^2 + 
	\frac{\sigma^2_{\max}(\Ab_s)}{\sigma^2_{\min}(\Ab_l)} \nt{\deltab_s}^2 
	= \left( 
	1 + \frac{\sigma^2_{\max}(\Ab_s)}{\sigma^2_{\min}(\Ab_l)} 
	\right)
	\nt{\deltab_s}^2\cdot
	\label{eq: prop tight up delta}
\end{equation}
Finally, we use (\ref{eq: unconst max norm y}) again to write $\nt{\deltab_s} \le \alpha \sqrt{m_s}$, which in combination with the above inequality gives (\ref{eq: tight prop}).
\end{IEEEproof}

\medskip

To prove Theorem~\ref{th: gamma bound}, we also need the following lemma, proof of which is left to Appendix~\ref{app: gamma1 lt gamma0}.
\begin{lemma}\label{lem: gamma1 lt gamma0}
Let $\Ab$ be an $n \times m$ ($m>n$) matrix with unit $\ell^2$ norm columns. Then $\gamma_1(\Ab) \ge \sqrt{m}\,$.
\end{lemma}

\medskip

\begin{IEEEproof}[Proof of Theorem~\ref{th: gamma bound}]
Note that $(\hsbb-\sbb_0)\in\mathrm{null}(\Ab)$. Moreover, $\sbb_0$ has at most $\lfloor \frac{\ell}{2} \rfloor$ nonzero components and $\hsbb$ has at most $\lfloor \frac{\ell}{2} \rfloor$ components with magnitudes larger than $\alpha_{\hsbb, \ell}$. Therefore, $\hsbb-\sbb_0$ has at most $\ell$ components with magnitudes larger than $\alpha_{\hsbb, \ell}$, that is, it has either 0, or 1, \dots, or $\ell$ components with magnitudes larger than $\alpha_{\hsbb, \ell}$. If it has $1 \le j \le \ell$ components larger than $\alpha_{\hsbb, \ell}$, from (\ref{eq: tight prop}) we have
\begin{equation}
\nt{\hsbb-\sbb_0} \le 
 \alpha_{\hsbb, \ell}
\sqrt{(m-j) \left( 1 + \frac{\sigma^2_{\max}(\Ab_s)}{\sigma^2_{\min}(\Ab_l)}\right)}
  \le \gamma_j(\Ab) \alpha_{\hsbb, \ell},
\end{equation}
because $\gamma_j$ had been defined as the maximum of $\sqrt{(m-j) \left( 1 + {\sigma^2_{\max}(\Ab_s)}/{\sigma^2_{\min}(\Ab_l)}\right)}$ for all possible choices of $\Ab_l$ and $\Ab_s$. On the other hand, if $\hsbb-\sbb_0$ has no components larger than $\alpha_{\hsbb, \ell}$, from (\ref{eq: tight lemma a})  we have $\nt{\hsbb-\sbb_0} \le \alpha_{\hsbb, \ell} \sqrt{m}$. This completes the proof of (\ref{eq: tight bound nonnorm}). Then, combining it with Lemma~\ref{lem: gamma1 lt gamma0} proves (\ref{eq: tight bound}).
\end{IEEEproof}

\medskip

\noindent {\bf Experiment. }
To experimentally compare the ``first bound'' given in (\ref{eq: sigmin bound}) and the ``second bound'' given in (\ref{eq: tight bound}), we conduced a simple experiment. We firstly generated a random $\Ab$ of dimension $8\times 12$ by generating each of its entries using $N(0,1)$ distribution, and then divided each of its columns by its norm to obtain a unit norm column matrix $\Ab$. We generated then a random sparse $\sbb_0$ by randomly choosing the positions of $p=2$ of its entries and then assigning random magnitudes (drawn from a $N(0,1)$ distribution) to these positions and zero to other positions. Then $\xb=\Ab\sbb_0$ was calculated and $\xb$ and $\Ab$ were given to SL0 algorithm and the parameter $\sigma_{\min}$ of SL0 was chosen relatively large ($0.1$) to force SL0 to create a not so accurate estimation $\hsbb$. Then the actual error $\nt{\hsbb-\sbb_0}$, the ``first bound'' and the ``second bound'' were calculated by setting $\ell=2p$ (note that $\Ab$ is of relatively small dimensions, permitting exact calculation of $\varsm{\ell}{\Ab}$ and $\bar\gamma(\Ab)$ using a combinatorial search). The whole experiment was repeated 100 times by regenerating $\Ab$ and $\sbb_0$. The average values of the ratios (First bound)/(Actual error) and (Second bound)/(Actual error) through these 100 experiments were 47.4 and 16.2, respectively. It is seen that the second bound is highly tighter than the first bound. Moreover, although the ratio (Second bound)/(Actual error) is seen to be in average 16.2, this bound is in fact a tight bound, in the sense that there are instances of $\Ab$, $\sbb_0$ and $\hsbb$ such that the equality in (\ref{eq: tight bound}) holds. Such an example is given in the next subsection, proving the tightness of this bound.

\subsection{Example of equality in Theorem~\ref{th: gamma bound}} \label{sec: Tight Example}
To show that the bound given in Theorem~\ref{th: gamma bound} is tight, we present the following tricky example, which is stated in the form of a proposition.

\begin{prop}\label{prop: example}
The estimation error $\nt{\hsbb - \sbb_0}$ achieves its upper bound in (\ref{eq: tight bound URP}) for
\begin{equation}
	\Ab = \left[
	\begin{array}{ccc}
	   1 & \cos \theta & \sin \frac{\theta}{2} \\
	   0 & \sin \theta & - \cos \frac{\theta}{2}
	\end{array}
	\right], \,
	\sbb_0 = \left[
	\begin{array}{c}
	   \beta \\
	   0 \\
	   0
	\end{array}
	\right], \,
	\hsbb = \left[
	\begin{array}{c}
	   0 \\
	   \beta \\
	   \alpha
	\end{array}
	\right],
   \label{eq: A tight example} 
\end{equation}
for any $0<\theta< \cos^{-1}(\frac{\sqrt{17}-1}{4})\approx 38.6683^\circ$, any $\alpha>0$, and $\beta\triangleq\alpha/(2 \sin \frac{\theta}{2})$.
\end{prop}

For example, for $\theta=5^\circ$ and $\alpha=0.2$, we have the following example (up to 4 digits) which achieves the upper bound of (\ref{eq: tight bound URP}):
\begin{equation*}
	\Ab = \left[\!\!\!
	\begin{array}{ccc}
	   1 & 0.9962 & 0.0436 \\
	   0 & 0.0872 & -.9990
	\end{array}\!\!\!
	\right], \,
	\sbb_0 \!=\! \left[\!\!\!
	\begin{array}{c}
	   2.2926 \\
	   0 \\
	   0
	\end{array}\!\!\!
	\right], \,
	\hsbb \!=\! \left[\!\!\!
	\begin{array}{c}
	   0 \\
	   2.2926 \\
	   0.2
	\end{array}\!\!\!
	\right].
\end{equation*}

Note that $\hsbb$ in this example is an approximately sparse solution of $\Ab \sbb = \xb$, where $\xb\triangleq \Ab \sbb_0$, but it is completely different from $\sbb_0$.

For proof, we need the following lemma:
\begin{lemma} \label{lem: sv of 1 and 2 col}
a) Let $\Bb$ be a single-column matrix (\ie\ a column vector), and this column has unit $\ell^2$ norm. Then the sole singular value of $\Bb$ is equal to 1.

b) Let $\Bb$ be two-column matrix (with more than one row), columns of which, $\bb_1$ and $\bb_2$, have unit $\ell^2$ norm. Then the two singular values of $\Bb$ are equal to $\sigma_{\min}^2(\Bb)=1-\rho$ and $\sigma_{\max}^2(\Bb)=1+\rho$, where $\rho\triangleq |\bb_1^T \bb_2| = |\cos\varphi|$, in which $\varphi$ is the angle between $\bb_1$ and $\bb_2$. Note that a smaller angle $\varphi$, results both in smaller $\sigma_{\min}$ and larger $\sigma_{\max}$.
\end{lemma}
\begin{IEEEproof}
  The result is simply obtained by direct calculations of the eigenvalues of $\Bb^T\Bb$.
\end{IEEEproof}

\begin{IEEEproof}[Proof of Proposition~\ref{prop: example}]
{\em Step 1)} Note that $\Ab$ satisfies the URP. By defining $\xb \triangleq \Ab \sbb_0$, it can be easily verified that both $\sbb_0$ and $\hsbb$ are solutions of $\Ab \sbb = \xb$. Moreover, for $0<\theta<60^\circ$, we have $\beta>\alpha$ and hence $\alpha_{\hsbb, n}=\alpha$.

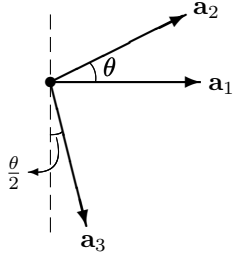
\begin{figure}[tb]
\setlength{\unitlength}{2mm}%
\centering

\begin{picture}(19,16)(-7,-10)
 \thicklines
 \put(0,0){\vector(1,0){10}}
 \put(10.5,-0.5){$\ab_1$}

 \put(0,0){\vector(2,1){9}}
 \put(9.5,4.5){$\ab_2$}

 \put(0,0){\vector(1,-4){2.4}}
 \put(2,-11){$\ab_3$}

 \put(0,0){\circle*{0.7}}

 \thinlines
 \multiput(0,-10)(0,1.5){10}{\line(0,1){1}}

 \qbezier(3,0)(3.1,0.8)(2.6,1.3)
 \put(3.5,0.4){$\theta$}

 \qbezier(0,-3.5)(0.6,-3.5)(0.8,-3.3)
 \qbezier(0.4,-3.7)(1,-6)(-1,-6)
 \put(-1,-6){\vector(-1,0){0.5}}
 \put(-3,-6.5){$\frac{\theta}{2}$}
 \put(3.5,0.4){$\theta$}

\end{picture}

\caption{The columns of the matrix $\Ab$ in (\ref{eq: A tight example}) are shown as the vectors $\ab_1$, $\ab_2$ and $\ab_3$.}
\label{fig: Tight Example}

\end{figure}

{\em Step 2) Calculating $\gamma_1(\Ab)$}: From definition (\ref{eq: eta def}), for $j=1$, $\Bb$ has only one column and hence $\sigma_{\min}(\Bb)=1$ from Lemma~\ref{lem: sv of 1 and 2 col}. Let $\ab_1$, $\ab_2$ and $\ab_3$ denote the columns of $\Ab$, respectively. These vectors are drawn in Fig.~\ref{fig: Tight Example}. To maximize $\sigma_{\max}(\Bbbar)$ among the 3 possible choices for $\Bbbar$, using Lemma~\ref{lem: sv of 1 and 2 col}, we have to find the two vectors for which the absolute value of the cosine of their angles is maximum. Simple manipulation of Fig.~\ref{fig: Tight Example} shows that for $0<\theta< 60^\circ$ the maximum of $\sigma_{\max}(\Bbbar)$ is obtained for $\Bbbar=[\ab_1, \ab_2]$. Consequently, $\eta^2_1(\Ab)=1+\cos\theta \Rightarrow \gamma_1^2(\Ab)=2 (2+ \cos\theta)$.

{\em Step 3) Calculating $\gamma_2(\Ab)$}: From definition (\ref{eq: eta def}), for $j=2$, $\Bbbar$ has only one column and hence $\sigma_{\max}(\Bbbar)=1$ from Lemma~\ref{lem: sv of 1 and 2 col}. Among the 3 possible choices for $\Bb$, using Lemma~\ref{lem: sv of 1 and 2 col}, it can be seen that for $0<\theta< 60^\circ$ the minimum of $\sigma_{\min}(\Bb)$ is obtained for $\Bb=[\ab_1, \ab_2]$. Consequently, $\eta_2^2(\Ab)=1/(1-\cos\theta)\Rightarrow \gamma_2^2(\Ab)= 1+1/(1-\cos\theta)$.

{\em Step 4) Comparing $\gamma_1(\Ab)$ and $\gamma_2(\Ab)$}: Simple algebra shows that
\begin{gather*}
  \gamma_2^2(\Ab)>\gamma_1^2(\Ab) \Leftrightarrow 
  2\cos^2\theta+\cos\theta-2>0 \Leftrightarrow \\
  \cos\theta>\frac{\sqrt{17}-1}{4}  \Leftrightarrow  
  |\theta| < \theta_0,
\end{gather*}
where $\theta_0\triangleq\cos^{-1}(\frac{\sqrt{17}-1}{4})\approx 38.6683^\circ$. Hence, since by the assumption $0<\theta<\theta_0$, we will have
\begin{equation}
	\bar\gamma(\Ab)=\gamma_2(\Ab)=\sqrt{1+\frac{1}{1-\cos\theta}}=\sqrt{1+\frac{1}{2 \sin^2\frac{\theta}{2}}}\,\cdot
\end{equation}

{\em Step 5)} Now, for $\beta=\alpha/(2 \sin \frac{\theta}{2})$ we write
\begin{equation*}
\nt{\hsbb-\sbb_0}^2=2 \beta^2 + \alpha^2=\alpha^2\left(1+\frac{1}{2\sin^2\frac{\theta}{2}}\right)=\bar\gamma^2(\Ab)\alpha^2,
\end{equation*}
which completes the proof.
\end{IEEEproof}

\section{The noisy case}\label{sec: noisy}
Instead of the noiseless system (\ref{eq: noiseless system}), consider now the noisy case
\begin{equation}
	\xb = \Ab \sbb + \nb,
	\label{eq: noisy system}
\end{equation}
where $\nb$ denotes the noise vector, and $\nt{\nb}\le \epsilon$ with $\epsilon\ge 0$. In SCA applications, $\nb$ denotes the measurement noise in sensors, and in sparse signal decomposition applications, (\ref{eq: noisy system}) is for modeling approximate signal decomposition, in which $\epsilon$ is the acceptable tolerance of the decomposition.

In presence of noise, the minimum $\ell^0$ norm solution of $\Ab \sbb =\xb$ is not stable, in the sense that if $\xb\triangleq \Ab \sbb_0+\nb$ (in which $\sbb_0$ is sparse), then the minimum $\ell^0$ norm solution of $\Ab \sbb=\xb$ may be completely different from $\sbb_0$, even for very small amount of noise~\cite{DonoET06,Wohl03}. Hence, instead of finding the sparsest solution of $\xb=\Ab\sbb$, it is proposed to estimate $\sbb_0$ as~\cite{DonoET06}
\begin{equation}
	\hsbb = \argmin_{\sbb} \nz{\sbb} \quad \mbox{s.t.} \quad \nt{\Ab \sbb-\xb}\le\delta,
	\label{eq: P 0 delta}
\end{equation}
for a $\delta \ge \epsilon \ge 0$. This approach insures the stability in the sense that $\nt{\hsbb - \sbb_0}$ grows at worst proportionally to the noise level~\cite{DonoET06,BabaJ09} (other variants of the above expression have also been studied in the literature, for example replacing the $\ell^0$ norm with $\ell^1$ norm, or replacing the $\ell^2$ norm in $\nt{\Ab \sbb-\xb}$ by $\ell^1$ and $\ell^\infty$ norms~\cite{DonoET06,Fuch05,Trop06elsevier,Trop06,Trop09}).

In this section, we are going to study the generalization of the main question of previous sections to this noisy case:
If we have an estimation $\hsbb$ satisfying $\nt{\Ab \hsbb-\xb}\le\delta$, and if $\hsbb$ is sparse only in the approximate sense\footnote{Such a solution may be obtained for example using Robust-SL0~\cite{EfteBJA09}.}, is it possible (without knowing $\sbb_0$) to construct an upper bound on the error $\nt{\hsbb - \sbb_0}$? Indeed, we will firstly generalize the looser bound (\ref{eq: sigmin bound}). Then, it will be seen that the tight bound (\ref{eq: tight bound}) would not be easy to generalize as a closed form formula.

\subsection{Generalizing the looser bound (\ref{eq: sigmin bound})}

\subsubsection{The theorem}
The generalization of the looser bound to the noisy case is given by the following theorem:

\begin{theorem}\label{th: sigmin bound noisy}
Let $\Ab$ be an ${n \times m}$ matrix ($m>n$) with unit $\ell^2$ norm columns.
Let $\xb = \Ab \sbb_0 + \nb$, where $\nt{\nb}\le \epsilon$ and $\|\sbb_0\|_0 \le {\ell}/{2}$, in which $\ell$ is an arbitrary integer less than or equal to $q(\Ab)$. Let $\hsbb$ be an estimation of $\sbb_0$ satisfying $\nt{\Ab \hsbb - \xb}\le \delta$, and define $\alpha_{\hsbb,\ell}\triangleq h(\lfloor {\ell}/{2} \rfloor+1,\hsbb)$. Then
\begin{equation}
  \label{eq: sigmin bound noisy}
	\|\hsbb-\sbb_0\|_2 \le \left( \frac{1}{\varsm{\ell}{\Ab}} + 1 \right) m \alpha_{\hsbb,\ell} + \frac{\Delta}{\varsm{\ell}{\Ab}},
\end{equation}
where $\Delta\triangleq\epsilon+\delta$.
\end{theorem}

\medskip

\noindent {\bf Remark 1. } 
For $\Delta=0$ (which corresponds to the noiseless case), the above bound reduces to the looser bound for the noiseless case, \ie\ (\ref{eq: sigmin bound noisy}) reduces to (\ref{eq: sigmin bound}).

\noindent {\bf Remark 2. } 
It is also interesting to consider the case $\alpha_{\hsbb,\ell}=0$. It corresponds to the case where $\hsbb$ is sparse in the exact sense, \eg\ where $\hsbb$ is an exact solution of (\ref{eq: P 0 delta}) for a $\delta \ge \epsilon$. In this case, the bound (\ref{eq: sigmin bound noisy}) becomes
\begin{equation}
	\|\hsbb-\sbb_0\|_2 \le \frac{\epsilon+\delta}{\varsm{\ell}{\Ab}}\cdot
	\label{eq: stability whole range}
\end{equation}
The above inequality holds for all values of $\ell$ satisfying the conditions of the theorem, and hence also for  $\ell=q(\Ab)$ (which is the largest possible $\ell$). This proves that {\em the problem (\ref{eq: P 0 delta}) is stable for all $\nz{\sbb_0}<\frac{1}{2}\spark(\Ab)$, \ie\ for the whole range of uniqueness of the sparse solution}, while in~\cite{DonoET06}, this stability had been proved only for the highly more limited range $\nz{\sbb_0}<\frac{1+M^{-1}}{2}$, where $M$ is the mutual coherence of $\Ab$. We have discussed this generalization of the stability and the inequality (\ref{eq: stability whole range}) in the correspondence\footnote{In fact, in that correspondence, we have even presented a more general result: we have shown that even minimizing $\nz{\sbb}$ in (\ref{eq: P 0 delta}) is not necessary, that is, the stability is resulted only from $\nt{\Ab \sbb - \xb}\le \delta$ for the whole uniqueness range $\nz{\sbb_0}<\frac{1}{2}\spark(\Ab)$, provided that the estimation $\hsbb$ satisfies also $\nz{\hsbb}<\frac{1}{2}\spark(\Ab)$.}~\cite{BabaJ09}.

\subsubsection{Proof}
To prove the above theorem, we need to first generalize Proposition~\ref{prop: sigmin bound} to the noisy case. This is given in the following Proposition, which is based on a modification of~\cite[Lemma~4]{MohiBJ09}:
\begin{prop}\label{prop: sigmin delta noisy}
Let $\Ab$ be an ${n \times m}$ matrix ($m>n$)  with unit $\ell^2$ norm columns, and assume
that every $\ell$ columns of $\Ab$ are linearly independent ($\ell \le n$). Let $\deltab$ be a vector satisfying $\nt{\Ab \deltab}\le \Delta$ for a constant $\Delta\ge 0$. If for $\alpha\ge 0$, $\|\deltab\|_{0,\alpha} \le \ell$, then
\begin{equation}\label{eq: sigmin delta noisy}
\nt{\deltab} \le \left( \frac{1}{\varsm{\ell}{\Ab}} + 1 \right) m \alpha + \frac{\Delta}{\varsm{\ell}{\Ab}}\cdot
\end{equation}
\end{prop}

\begin{IEEEproof}
The proof is based on modifications in the proof of Proposition~\ref{prop: sigmin bound}. Let $\deltab_l$, $\deltab_s$ be as defined in that proof.

{\bf Case 1} ($m_l=0$ and $m_s=m$): Exactly like the proof of Proposition~\ref{prop: sigmin bound}, $\|\deltab\|_2  \le m\alpha$ which satisfies also (\ref{eq: sigmin delta noisy}).

{\bf Case 2} ($1 \le m_l\le \ell$ and  $m_s\le m-1$): Let $\Ab_l$ and $\Ab_s$ be as defined in the proof of  Proposition~\ref{prop: sigmin bound}. We define again $\bb\triangleq\Ab_l\deltab_l$, but $\bb$ is no more equal to $-\Ab_s\deltab_s$. Similar to (\ref{eq: prop nontight up b}) we have $\nt{\Ab_s \deltab_s}\le m \alpha$. For upper bounding $\nt{\bb}$, instead of (\ref{eq: prop nontight up b}), we use the general inequality $\nt{\yb_1}-\nt{\yb_2}\le \nt {\yb_1+\yb_2}$ to write
\begin{gather}
	\underbrace{\nt{\Ab_l \deltab_l}}_{\nt{\bb}} - \nt{\Ab_s \deltab_s} \le \nt{\Ab_l \deltab_l + \Ab_s \deltab_s} \le \Delta \nonumber \\
	\Rightarrow \nt{\bb} \le \nt{\Ab_s \deltab_s} + \Delta \le m \alpha + \Delta \label{eq: prop nontight up b noisy}.
\end{gather}
Putting this in (\ref{eq: prop nontight up d_l based on b}), we have (instead of (\ref{eq: prop nontight up d_l}))
\begin{equation}
	\|\deltab_l\|_2 \le \frac{m\alpha+\Delta}{\sigma_{\min}(\Ab_l)} \label{eq: prop nontight up d_l noisy},
\end{equation}
and hence (\ref{eq: prop nontight up delta}) becomes
\begin{equation}
	\|\deltab\|_2 \le \|\deltab_l\|_2 + \|\deltab_s\|_2 \le \frac{m\alpha+\Delta}{\sigma_{\min}(\Ab_l)} + m \alpha,
	\label{eq: prop nontight up delta noisy}
\end{equation}
which in combination with $\sigma_{\min}(\Ab_l)\ge \varsm{\ell}{\Ab}$ completes the proof.
\end{IEEEproof}

\medskip

\begin{IEEEproof}[Proof of Theorem~\ref{th: sigmin bound noisy}]
Similar to the proof of Theorem~\ref{th: sigmin bound}, $\hsbb-\sbb_0$ has at most $\ell$ components with magnitudes larger than $\alpha$. However, here $\hsbb-\sbb_0$ is not necessarily in $\mbox{null}(\Ab)$. Instead, we use the general inequality $\nt{\yb_1}-\nt{\yb_2}\le \nt {\yb_1-\yb_2}$ to write
\begin{gather}
	\nt{\Ab(\hsbb-\sbb_0)} - \nt{\nb} \le \nt{\Ab\hsbb-\Ab\sbb_0-\nb}=\nt{\Ab\hsbb-\xb}\le\delta \nonumber \\
	\Rightarrow \nt{\Ab(\hsbb-\sbb_0)} \le \nt{\nb}+\delta \le \epsilon+\delta.
\end{gather}
Therefore, $\deltab\triangleq\hsbb-\sbb_0$ satisfies the conditions of Proposition~\ref{prop: sigmin delta noisy} for $\Delta=\epsilon+\delta$, which completes the proof.
\end{IEEEproof}

\subsection{Generalizing the tight bound (\ref{eq: tight bound})}
In this section, we will see that obtaining a `closed form' expression as a generalization of the tight bound (\ref{eq: tight bound}) to the noisy case would not be easy. In fact, it is seen that the generalization of the looser bound was based on generalizing Proposition~\ref{prop: sigmin bound} to Proposition~\ref{prop: sigmin delta noisy}.
Similarly, we can also generalize Proposition~\ref{prop: sigmin delta noisy} to the noisy case:
\begin{prop}\label{prop: tight noisy}
Let $\Ab$, $\ell$, $\deltab$, $\alpha$ and $\Delta$ be as in Proposition~\ref{prop: sigmin delta noisy}.
Let $m_l$ and $m_s$ denote the number of entries of $\deltab$ with magnitudes larger than, and less than or equal to $\alpha$, respectively.

a) Case $m_l=0$ and $m_s=m$: We have
\begin{equation}\label{eq: tight lemma a noisy}
\|\deltab\|_2 \le \sqrt{m} \, \alpha.
\end{equation}

b) Case $1 \le m_l\le \ell$ and  $m_s\le m-1$: 
Let $\Ab_l$ and $\Ab_s$ be composed of the columns of $\Ab$ which correspond to the entries of $\deltab$ that have magnitudes larger than $\alpha$, and less than or equal $\alpha$, respectively. Then
\ifonecol 
\begin{equation}\label{eq: tight lemma noisy}
\nt{\deltab}^2 \le \left( 1 + \frac{\sigma^2_{\max}(\Ab_s)}{\sigma^2_{\min}(\Ab_l)}\right)
m_s \alpha^2 + 2 \frac{\sigma_{\max}(\Ab_s)}{\sigma_{\min}(\Ab_l)} \sqrt{m_s} \, \alpha \Delta +
\frac{\Delta^2}{\sigma^2_{\min}(\Ab_l)}\cdot
\end{equation}
\else    
\begin{equation}\label{eq: tight lemma noisy}
\begin{split}
\nt{\deltab}^2 \le & \left( 1 + \frac{\sigma^2_{\max}(\Ab_s)}{\sigma^2_{\min}(\Ab_l)}\right)
 m_s \alpha^2 + \\
 & + 2 \frac{\sigma_{\max}(\Ab_s)}{\sigma_{\min}(\Ab_l)} \sqrt{m_s} \, \alpha \Delta +
 \frac{\Delta^2}{\sigma^2_{\min}(\Ab_l)}\cdot
\end{split}
\end{equation}
\fi   
\end{prop}

\begin{IEEEproof}
Part (a): The proof is the same as the proof of part (a) of Proposition~\ref{prop: tight}. 

Part (b): Let again $\bb\triangleq\Ab_l\deltab_l$. With similar reasoning as in (\ref{eq: prop nontight up b noisy}) we have $\nt{\bb} \le \nt{\Ab_s \deltab_s} + \Delta$. Moreover, $\|\Ab_s \deltab_s\|_2 \le \sigma_{\max}(\Ab_s) \nt{\deltab_s}$, and hence
\begin{equation}
 \nt{\bb} \le \sigma_{\max}(\Ab_s)\nt{\deltab_s} + \Delta 
 \label{eq: prop tight up b noisy}.
\end{equation}
Combining it with (\ref{eq: prop nontight up d_l based on b}), we will have
\begin{equation}
	\|\deltab_l\|_2 \le \frac{\sigma_{\max}(\Ab_s)\nt{\deltab_s}+\Delta}{\sigma_{\min}(\Ab_l)}\cdot
	\label{eq: prop tight up d_l noisy}
\end{equation}
Moreover, $\nt{\deltab_s}\le \sqrt{m_s}\, \alpha$ (from Eq.~(\ref{eq: unconst max norm y})). Combining this, the above inequality, and $\nt{\deltab}^2=\nt{\deltab_s}^2+\nt{\deltab_l}^2$ proves the proposition.
\end{IEEEproof}

\medskip 

However, although Proposition~\ref{prop: sigmin delta noisy} was generalized to Proposition~\ref{prop: tight noisy}, it would be tricky to use it to obtain a closed form generalization of (\ref{eq: tight bound}) to the noisy case. To see the difference, recall the argument of using (\ref{eq: tight prop}) to obtain (\ref{eq: tight bound}) (which is the same argument as using (\ref{eq: prop sigmin}) to obtain (\ref{eq: sigmin bound})): under the conditions of Theorem~\ref{th: gamma bound}, $\deltab\triangleq\hsbb-\sbb_0$ satisfies the conditions of Proposition~\ref{prop: tight} for $\alpha=\alpha_{\hsbb,\ell}$. However, since we don't know $\sbb_0$, we don't know which components of $\deltab$ are smaller and which ones are larger than $\alpha$, and hence, we don't know $\Ab_s$ and $\Ab_l$. Therefore, we consider the worst case of the bound given by (\ref{eq: tight prop}), that is, we maximize the right hand side of (\ref{eq: tight prop}) on all possible partitionings of $\Ab$ into $\Ab_l$ and $\Ab_s$ (where the number of columns of $\Ab_l$ is at most equal to $\ell$). The point is that the right hand side of (\ref{eq: tight prop}) was in a form that its maximization with respect to all possible partitionings of $\Ab$ was independent of $\alpha$, and gave us the bound (\ref{eq: tight bound}), in which, we had a constant $\bar\gamma(\Ab)$ which is independent of $\alpha$, and depends only on the dictionary.

However, with the same reasoning, to obtain an upper bound on $\nt{\hsbb-\sbb_0}$ under the conditions of Theorem~\ref{th: sigmin bound noisy}, we have to maximize the right hand side of (\ref{eq: tight lemma noisy}) with respect to all possible partitionings of $\Ab$ into $\Ab_l$ and $\Ab_s$ (where the number of columns of $\Ab_l$ is at most equal to $\ell$). However, here, this maximization depends also on $\alpha$ and $\Delta$, because $\sigma_{\max}(\Ab_s)/\sigma_{\min}(\Ab_l)$ and $1/\sigma_{\min}(\Ab_l)$ are not necessarily maximized\footnote{Using a small MATLAB code, it is easy to find examples of $\Ab$ for which these two quantities are maximized for different partitionings.} for the same partitioning of $\Ab$. Consequently, we can probably say nothing better than:

\begin{theorem}\label{th: gamma bound noisy}
Let all parameters be as defined in Theorem~\ref{th: sigmin bound}. Let $\Bb\in \Pc_j(\Ab)$ and let $\Bbbar$ denote the matrix composed by the columns of $\Ab$ that are not in $\Bb$. Define
\ifonecol
\begin{equation}
  f(\Ab, \alpha, \Delta) \triangleq \max_{1\le j \le \ell} \, \, \max_{\Bb\in\Pc_j(\Ab)} \left\{
	\left( 1 + \frac{\sigma^2_{\max}(\Bbbar)}{\sigma^2_{\min}(\Bb)}\right)
	m_s \alpha^2 + \right. 
	\left. + 2 \frac{\sigma_{\max}(\Bbbar)}{\sigma_{\min}(\Bb)} \sqrt{m_s} \, \alpha \Delta +
	\frac{\Delta^2}{\sigma^2_{\min}(\Bb)} \right\}\cdot
\end{equation}
\else
\begin{align}
  f(\Ab, \alpha, \Delta) \triangleq & \max_{1\le j \le \ell} \, \, \max_{\Bb\in\Pc_j(\Ab)} \left\{
	\left( 1 + \frac{\sigma^2_{\max}(\Bbbar)}{\sigma^2_{\min}(\Bb)}\right)
	m_s \alpha^2 + \right. \nonumber \\
	& \left. + 2 \frac{\sigma_{\max}(\Bbbar)}{\sigma_{\min}(\Bb)} \sqrt{m_s} \, \alpha \Delta +
	\frac{\Delta^2}{\sigma^2_{\min}(\Bb)} \right\}\cdot
\end{align}
\fi
Then
\begin{equation}
	\|\hsbb-\sbb_0\|_2^2 \le \max \la m \alpha_{\hsbb,\ell}^2\, ,  f(\Ab, \alpha_{\hsbb,\ell}, \Delta) \ra\cdot
\end{equation}
Moreover, if $\Ab$ has unit $\ell^2$ norm columns, then
\begin{equation}
	\|\hsbb-\sbb_0\|_2^2 \le f(\Ab, \alpha_{\hsbb,\ell}, \Delta).
	\label{eq: tight bound noisy}
\end{equation}
\end{theorem}

\smallskip

\noindent {\bf Remark. } 
In Theorems~\ref{th: sigmin bound} to~\ref{th: sigmin bound noisy}, the quantities $\bar\gamma(\Ab)$ or $\varsm{n}{\Ab}$ are calculated only once for each dictionary, and then they are used with $\alpha_{\hsbb,\ell}$ (and probably $\Delta$) of a specific problem. However, in the above theorem, the dictionary, $\alpha$ and $\Delta$ interact, and hence the upper bound should be calculated for each specific problem separately  and since this calculation is NP-hard, its usage in practical problems is probably limited.

\section{Random Dictionaries}\label{sec: random}
Theorems~\ref{th: sigmin bound} and~\ref{th: gamma bound} suggest that $\sm{\ell}{\Ab}$ and/or $\bar\gamma_\ell(\Ab)$ are important parameters of a dictionary. However, estimating these parameters for 
a deterministic matrix seems to be NP-hard (this has already been proven for estimating $\sm{\ell}{\Ab}$~\cite{CivrM09}). In effect, calculating $\sm{\ell}{\Ab}$ requires examination of all $n\times \ell$ submatrices of $\Ab$, and calculation of $\bar\gamma_\ell(\Ab)$ requires examination of all $n\times j$ submatrices of $\Ab$ for $j=1,\dots,\ell$. These tasks are combinatorial and intractable (although they have to be done only once for a given dictionary). Moreover, even finding a computationally tractable lower bound for $\sm{q}{\Ab}$ or an upper bound for $\bar\gamma'(\Ab)$ would provide us a computable upper bound for the error.

On the contrary, for a random $\Ab$ with independent and identically distributed (iid) entries, we need no more to examine all of its $\binom{n}{\ell}$ submatrices to obtain probabilistic upper bounds, because all $n \times \ell$ submatrices are statistically identical. On the other hand, singular values of random matrices have extensively been studied in the literature~\cite{Edel89,TuliV04}. Indeed, it is well-known that the singular values of random matrices are not ``so random'' and are highly concentrated around some deterministic values~\cite[Th. 2.7]{DaviS01}. Random dictionaries are also practically important, \eg\ they are frequently used in compressed sensing~\cite{Bara07}. Note also that random matrices satisfy the URP with probability 1.

In this section, we consider random dictionaries, and state some probabilistic upper bounds for the estimation error $\nt{\hsbb - \sbb_0}$ without knowing $\sbb_0$, and independent of the method used for estimating $\hsbb$.

\subsection{Review of some results from random matrix theory}\label{sec: random matrix review}
Let $\Xb$ be an $n \times p$ random matrix with independent and identically distributed (iid) entries with zero mean and variance $\frac{1}{n}$ (hence the expected values of the $\ell^2$ norm of its columns are equal to 1).
A famous result by Mar\u chenko and Pastur~\cite[Th. 2.35]{TuliV04} states that if the entries of $\Xb$ come from any distribution with fourth order moment of order $O(\frac{1}{n^2})$, as $n,p\to\infty$  and $\frac{p}{n}\to c>0$, the empirical distribution of singular values of $\Xb$ converges almost surely (a.s.) to a distribution bounded between $|1-\sqrt{c}\,|$ and $1+\sqrt{c}$. Moreover, if the entries come from any distribution with finite fourth order moment, $\sigma_{\max}(\Xb)$ has been shown~\cite{YinBK88} to converge a.s.~to $1+\sqrt{c}$ (this result has been firstly stated by Geman~\cite{Gema80} under some more restrictive conditions). Similarly, if $0<c<1$ (\ie\ for tall matrices), it has been shown (firstly in \cite{Silv85} for the Gaussian case and then in~\cite{BaiY93} for any distribution with finite fourth moment) that $\sigma_{\min}(\Xb)\xrightarrow{{a.s.}} 1-\sqrt{c}$. As it is said in~\cite{BaiY93}, it is obvious that a similar result holds also for wide matrices, that is, where $c>1$. To see this, let $c>1$. Then $\Xb'\triangleq \sqrt{{n}/{p}}\,\Xb^T$ is a tall $p\times n$ matrix with iid zero-mean elements with variance $\frac{1}{p}$, and hence, $\sqrt{{1}/{c}}\,\sigma_{\min}(\Xb)\xrightarrow{a.s.} 1-\sqrt{{1}/{c}}$, and consequently $\sigma_{\min}(\Xb)\xrightarrow{a.s.} \sqrt{c}-1$. Hence, generally, if $c\neq 1$, then $\sigma_{\min}(\Xb)\xrightarrow{a.s.} |1-\sqrt{c}\,|$. The case $c=1$ is, however, more complicated. For example, for  Gaussian square random matrices ($p=n$), if $n \to \infty$, then the probability density function (PDF) of the random variable $\sigma_{\min}(\Xb)$ converges to a simple known function~\cite[Th. 5.1]{Edel89},\,\cite{Shen01}. In other words, for $c\neq1$, as $n \to \infty$, the PDF of $\sigma_{\min}(\Xb)$ converges to a Dirac delta function (\ie\ $\sigma_{\min}(\Xb)$ converges a.s.~to a deterministic value), but this is not true for $c=1$.

Moreover, if $n\ge p$ (tall matrix), and the entries of $\Xb$ are drawn from a $N(0,1/n)$ distribution, then a result due to Davidson and Szarek~\cite[Th. 2.13]{DaviS01},\,\cite[Eqs. (4.35) and (4.36)]{CandT06} states that for any $r>0$
\begin{gather}
 \Prob \left\{ \sigma_{\max}(\Xb) > 1 + \sqrt{\frac{p}{n}} + r \right\} \le e^{-n r^2/2} \label{eq: Szarek bound sigmax},  \\
 \Prob \left\{ \sigma_{\min}(\Xb) < 1 - \sqrt{\frac{p}{n}} - r \right\} \le e^{-n r^2/2}. \label{eq: Szarek bound sigmin}
\end{gather} 
Note that the second inequality is mostly useful for $0<r<1- \sqrt{{p}/{n}}$ (otherwise it is trivial).

It is not difficult to see that similar equations hold also for wide matrices. To show it, let $\Xb$ be an $n \times p$ random matrix with $n <p$ (wide matrix) and with elements drawn from a $N(0,1/n)$ distribution (again normalized columns in expected value). Then, $\Yb\triangleq \sqrt{{n}/{p}}\,\Xb^T$ is a $p\times n$ tall matrix with $N(0,1/p)$ entries, for which we can write the above inequalities. For example, by writing (\ref{eq: Szarek bound sigmax}) for $\Yb$ we have
\begin{gather}
     \Prob \left\{ \sigma_{\max}(\Yb) > 1 + \sqrt{\frac{n}{p}} + r \right\} \le e^{-p r^2/2} \nonumber \\
     \Rightarrow\Prob \left\{ \sqrt{\frac{n}{p}} \sigma_{\max}(\Xb) > 1 + \sqrt{\frac{n}{p}} + r \right\} \le e^{-p r^2/2}  \nonumber \\
     \Rightarrow\Prob \left\{  \sigma_{\max}(\Xb) > \sqrt{\frac{p}{n}} + 1 + r\sqrt{\frac{p}{n}} \right\} \le e^{-p r^2/2}. \nonumber
\end{gather}
Hence, by defining $r' = r \sqrt{{p}/{n}}$, we have
\begin{equation*}
   \Prob \left\{  \sigma_{\max}(\Xb) > \sqrt{\frac{p}{n}} + 1 + r' \right\} \le e^{-p r'^2 (\frac{n}{p})/2} = e^{-n r'^2/2}.
\end{equation*}
Consequently, (\ref{eq: Szarek bound sigmax}) also holds for wide matrices. Similarly, it can be seen that for wide matrices ($p>n$), the inequality (\ref{eq: Szarek bound sigmin}) becomes
\begin{equation}
 \Prob \left\{ \sigma_{\min}(\Xb) < \sqrt{\frac{p}{n}} - 1 - r \right\} \le e^{-n r^2/2}.
\end{equation}

\subsection{Definitions and notations}
Let the dictionary $\Ab$ be a random $n\times m$ matrix with $m>n$ and with iid entries drawn from a $N(0,1/n)$ distribution. In this section, we use Theorem~\ref{th: gamma bound} and Davidson and Szarek inequalities to obtain a probabilistic upper bound for the error $\nt{\hsbb-\sbb_0}$. 

\noindent {\bf Remark.} 
Note that the $\ell^2$ norms of the columns of such an $\Ab$ are not necessarily equal to one (although their expected values are equal to one). Hence, we cannot use the bound given in Theorem~\ref{th: sigmin bound} for this $\Ab$, because that theorem requires that the columns of $\Ab$ have unit $\ell^2$ norms. Moreover, if we normalize the columns of $\Ab$ by dividing them by their $\ell^2$ norms, the new entries would no longer be independent, and consequently, we cannot use Davidson and Szarek inequalities and many other results in random matrix theory which require independent entries. Hence, it is not straightforward\footnote{In a personal email communication with the first author, Pr.~Szarek has generalized (\ref{eq: Szarek bound sigmin}) to the case where the elements of $\Ab$ are firstly drawn independently from a zero mean Gaussian distribution and then each column is divided by its $\ell^2$ norm to obtain a unit $\ell^2$ norm column dictionary. The final bound is looser than (\ref{eq: Szarek bound sigmin}) and is in a more complicated form. Hence obtaining a probabilistic bound based on Theorem~\ref{th: sigmin bound} is indeed possible, but is not straightforward. We don't consider it in this paper because the error bound is both looser and more complicated.} to obtain a probabilistic bound based on Theorem~\ref{th: sigmin bound} (this is a mistake that we had done in the conference paper~\cite{BabaMJ09}). However, Theorem~\ref{th: gamma bound} does not require unit $\ell^2$ norm columns, and we can use it to obtain a probabilistic upper bound on $\nt{\hsbb-\sbb_0}$.

\medskip

Note that the bound in Theorem~\ref{th: gamma bound} is based on the quantities $\eta_j(\Ab)$ and $\gamma_j(\Ab)$ defined in (\ref{eq: eta def}) and (\ref{eq: gamma def}), respectively. Hence to obtain a probabilistic bound on the error $\nt{\hsbb-\sbb_0}$, we obtain probabilistic upper bounds for these quantities.

For the random dictionary $\Ab$ defined above, for any division of $\Ab$ into $\Bb$ and $\Bb^c$ as stated in the definition of $\eta_j(\Ab)$, from the results of random matrix theory stated in the previous subsection, we expect that $\sigma_{\min}(\Bb)$ and $\sigma_{\max}(\Bb^c)$ are close to $1-\sqrt{j/n}$ and $1+\sqrt{(m-j)/n}$, respectively. Hence, $\eta_j(\Ab)$ and $\gamma_j(\Ab)$ are expected to be close to quantities
\begin{equation}
	\eta[j]  \triangleq \frac{1+\sqrt{\frac{m-j}{n}}}{1-\sqrt{\frac{j}{n}}}
	\label{eq: eta seq def}
\end{equation}
and 
\begin{equation}
	\gamma[j]  \triangleq \sqrt{(m-j)\left(1+\left(\eta[j]\right)^2\right)},
\end{equation}
respectively. More precisely, the results of the previous subsection imply that where $m,n\to\infty$ while $j/m$ converges to a constant and $j/n$ converges to a constant strictly smaller than 1, then $\eta_j(\Ab)$ and $\gamma_j(\Ab)$ will converge a.s.~to $\eta[j]$ and $\gamma[j]$, respectively.

To measure the deviation of $\eta_j(\Ab)$ and $\gamma_j(\Ab)$ from the above quantities (to larger values), let's define the shorthands
\begin{gather}
  \eta_{r_1r_2}[j] \triangleq \frac{1+\sqrt{\frac{m-j}{n}}+r_1}{1-\sqrt{\frac{j}{n}}-r_2}, \\
  \gamma_{r_1r_2}[j] \triangleq \sqrt{(m-j)\left(1+\left(\eta_{r_1r_2}[j]\right)^2\right)}\,. \label{eq: g r1r2 def}
\end{gather}

\smallskip

However, note that the bound of Theorem~\ref{th: gamma bound} depends mainly  on the quantities $\bar\gamma_\ell(\Ab)$ and $\bar\gamma'_\ell(\Ab)$, not $\gamma_j(\Ab)$. In other words, we need to maximize $\gamma[j]$ (or $\gamma_{r_1r_2}[j]$) over $j=1,2,\dots,\ell$. The following lemma, whose proof has been left to appendix, shows that the sequences defined above, \ie\ $\eta_{r_1r_2}[j]$ and $\gamma_{r_1r_2}[j]$ and hence $\eta[j]$ and $\gamma[j]$ (as the special case of $\eta_{r_1r_2}[j]$ and $\gamma_{r_1r_2}[j]$ for $r_1=r_2=0$), are all increasing with respect to $j$. Therefore, it shows that the above maximum is obtained for $j=\ell$.
\begin{lemma}\label{lem: g increasing}
  The sequence $\{\gamma_{r_1r_2}[j]\}$, $j=1,\dots,\ell$, where $\ell \le n-1$, is strictly increasing for all $r_1\ge0$ and $0\le r_2<1-\sqrt{\ell/n}$. Moreover, $\forall j: \gamma_{r_1r_2}[j] > \sqrt{m}$.
\end{lemma}

\smallskip

\noindent {\bf Remark 1. } 
The above lemma shows that the sequences $\eta[j]$ and $\eta_{r_1r_2}[j]$ are also increasing, because the product of $(1+\eta_{r_1r_2}^2[j])$ and the decreasing (and positive) sequence $\{m-j\}$ has become an increasing sequence.

\smallskip

\noindent {\bf Remark 2. } 
Note that for large dictionaries (more precisely where $m,n\to\infty$ while $j/m$ converges to a constant and $j/n$ converges to a constant strictly smaller than 1), $\gamma_j(\Ab)$ converges a.s.~to $\gamma[j]$. The above lemma states hence that for large dictionaries the sequence $\gamma_j(\Ab)$ is almost surely increasing and hence a.s.~$\bar\gamma_j(\Ab)=\gamma_j(\Ab)=\gamma[j]$. Moreover, the second part of the above lemma states that a.s.~$\bar\gamma'_j(\Ab)=\bar\gamma_j(\Ab)=\gamma_j(\Ab)=\gamma[j]$.

\smallskip

\noindent {\bf Remark 3. } 
In the Remark after Definition~\ref{defi: gamma} in Section~\ref{sec: eta defi}, we had provided an example of a matrix $\Ab$ for which the sequences $\eta_j(\Ab)$ and $\gamma_j(\Ab)$ were not increasing. The matrix $\Ab$ of that example was of very small size, and the above remark states that where the size of the dictionary grows finding such examples becomes more and more difficult.

\subsection{Probabilistic bounds on $\nt{\hsbb-\sbb_0}$}
In this section, we state two theorems as probabilistic upper bounds on the error, where the dictionary $\Ab$ is random with iid Gaussian entries. The first theorem states a bound for dictionaries of any size, whereas the second theorem considers the case of large dictionaries.

\begin{theorem}\label{th: Prob bound}
Let $\Ab$ be an $n \times m$, $m>n$, random matrix with iid and zero-mean Gaussian entries. Let $\ell$ be an integer in the range $1\le \ell \le n-1$.
Suppose that $\sbb_0$ is a solution of $\Ab \sbb = \xb$ for which $\|\sbb_0\|_0 \le {\ell}/{2}$. Let $\hsbb$ be a solution of $\Ab \sbb = \xb$, and define $\alpha_{\hsbb,\ell}\triangleq h(\lfloor {\ell}/{2} \rfloor+1,\hsbb)$. Then for all $r_1>0$ and $0<r_2<1-\sqrt{{\ell}/{n}}$\,,
\ifonecol
\begin{equation}
   \Prob \big\{ \nt{\hsbb-\sbb_0}>(\gamma_{r_1r_2}[\ell]) \alpha_{\hsbb,\ell} \big\} < 
     \left[ \binom{m}{1} + \binom{m}{2} + \cdots + \binom{m}{\ell}\right]
    \left( e^{-n r_1^2/2} + e^{-n r_2^2/2} \right).
 \label{eq: Th prob bound}
\end{equation}
\else
 \begin{multline}
    \Prob \big\{ \nt{\hsbb-\sbb_0}>(\gamma_{r_1r_2}[\ell]) \alpha_{\hsbb,\ell} \big\} < \\
     \left[ \binom{m}{1} + \binom{m}{2} + \cdots + \binom{m}{\ell}\right]
    \left( e^{-n r_1^2/2} + e^{-n r_2^2/2} \right).
 \label{eq: Th prob bound}
 \end{multline}
 
\fi
\end{theorem}

\noindent {\bf Remark 1. } 
When $\Ab$ is random as in Theorem~\ref{th: Prob bound}, it satisfies the URP with probability 1, and hence by the uniqueness theorem, any solution with $\nz{\sbb_0} \le \frac{n}{2}$ would be unique. Suppose, however, that $p \triangleq \|\sbb_0\|_0 < \frac{n}{2}$ (not $p\le\frac{n}{2}$) and set $\ell=2p$. Then $\ell \le n-1$ and hence from (\ref{eq: Th prob bound}),
\ifonecol
\begin{equation}
    \Prob \big\{ \nt{\hsbb-\sbb_0}>(\gamma_{r_1r_2}[2p]) \alpha_{\hsbb,2p} \big\} < 
    \left[ \binom{m}{1} + \binom{m}{2} + \cdots + \binom{m}{2p}\right]
    \left( e^{-n r_1^2/2} + e^{-n r_2^2/2} \right).
\label{eq: Th prob bound p}
\end{equation}
\else
\begin{multline}
    \Prob \big\{ \nt{\hsbb-\sbb_0}>(\gamma_{r_1r_2}[2p]) \alpha_{\hsbb,2p} \big\} < \\
     \left[ \binom{m}{1} + \binom{m}{2} + \cdots + \binom{m}{2p}\right]
    \left( e^{-n r_1^2/2} + e^{-n r_2^2/2} \right).
\label{eq: Th prob bound p}
\end{multline}
\fi

\noindent {\bf Remark 2: } 
Note that when $n$ grows, (\ref{eq: Th prob bound p}) does not necessarily provide a good upper bound on $\nt{\hsbb-\sbb_0}$, in the sense that as $n$ increases, the probability that $\nz{\hsbb-\sbb_0}>\gamma_{r_1r_2}[2p]\alpha_{\hsbb,2p}$ does not necessarily decreases exponentially to zero.
The point is that the maximum value for $r_2$ in Theorem~\ref{th: Prob bound} is $1-\sqrt{\ell/n}$, hence, where $n$ increases, although the term $n$ in $e^{-nr_2^2/2}$ increases, $r_2$ has to be smaller, and hence $e^{-nr_2^2/2}$ does not necessarily decrease.
In fact, the degree of sparsity of $\sbb_0$ plays an important role here.
For example, let choose $r_1=r_2$, and suppose that $2p$ is equal to its maximum theoretical value (which is $n-1$, because we had already excluded $p=n/2$ that turns $\gamma_{r_1r_2}[2p]$ to infinity). 
We expect heuristically that for large $n$'s, 
$\eta_{2p}(\Ab)$ is close to $\eta[2p]=(1-\sqrt{\frac{m-2p}{n}})/(1-\sqrt{\frac{2p}{n}})$. Taking into account the form of the denominator of $\eta_{r_1r_2}[2p]$, for having $\gamma_{r_1r_2}[2p]$ not too far from $\gamma[2p]$, let choose the value of $r_2$ as a small fraction of $1-\sqrt{{2p}/{n}}$, that is, let $r_2=c \cdot (1-\sqrt{{2p}/{n}})$, where $0<c<1$.  Then from (\ref{eq: Th prob bound p}),
\begin{equation}
    \Prob \big\{ \nt{\hsbb-\sbb_0}>(\gamma_{r_1r_2}[2p]) \alpha_{\hsbb,2p} \big\} < 
    A e^{-c^2 n (1-\sqrt{\frac{2p}{n}})^2/2}
    \label{eq: Th prob bound temp}
\end{equation}
where $A\triangleq2\sum_{j=1}^{2p}\binom{m}{j}$. Consider now the exponent $c^2 n (1-\sqrt{{2p}/{n}}\,)^2/2$. If $2p=n-1$, this exponent is in fact a {\em decreasing} function of $n$, and converges to 0 where $n\to\infty$. Consequently, by increasing $n$, not only $e^{-nr_2^2/2}$ does not decrease, but also it increases to 1. 

Another way to see the above problem is to note that, as stated after (\ref{eq: eta seq def}), for large matrices $\eta_{2p}(\Ab)$ converges to $\eta[2p]$ only if $2p/n$ converges to a value `strictly' smaller than 1. This is also seen from the discussion at the end of the first paragraph of Section~\ref{sec: random matrix review}.

On the other hand, if $p$ can be at most equal to a fraction of $n/2$, say $p=u \cdot (n/2)$, where $u<1$, then $e^{-nr_2^2/2}=\exp\la-c^2 n (1-\sqrt{u}\,)^2/2\ra$, which exponentially decreases where $n\to\infty$. The right hand side of (\ref{eq: Th prob bound p}) does not yet necessarily decrease, however, due to the combinatorial part. We can, however, state the following theorem, for smaller $u$'s (as will be discussed after the theorem):

\medskip

\begin{theorem}\label{th: random large dic}
Let $\Ab$ be an $n \times m$, $m>n$, random matrix with iid zero-mean Gaussian entries. Suppose that $\sbb_0$ is a solution of $\Ab \sbb = \xb$ with sparsity $p$, \ie , $p\triangleq\|\sbb_0\|_0$. 
Let $\hsbb$ be a solution of $\Ab \sbb = \xb$, and define $\alpha_{\hsbb,2p}\triangleq h(p+1,\hsbb)$. 
If $n\to\infty$, while $2p/n \to u<1$ and $m/2p\to v$, then for every $r_1>0$ and $0<r_2<1-\sqrt{u}$, 
with an exponentially increasing probability (with respect to $n$) we have
\begin{equation}
	\nt{\hsbb-\sbb_0} \le \gamma_{r_1r_2}[2p] \cdot \alpha_{\hsbb,2p}\, ,
	\label{eq: prob bound exponent}
\end{equation}
provided that
\begin{equation}
	u (1+ \ln v) < \min(r_1^2,r_2^2)/2.
	\label{eq: uv limitation}
\end{equation}
\end{theorem}

\medskip

\noindent {\bf Remark. } Condition (\ref{eq: uv limitation}) puts a limit on the maximum of the sparsity ($u$) for which the above theorem is applicable. To see this, let fix the underdeterminedness factor $\beta\triangleq m/n>1$. Then $v=\beta/u$, and hence (\ref{eq: uv limitation}) and $r_2<1-\sqrt{u}$ imply that
\begin{equation}
	u (1+\ln \frac{\beta}{u}) < (1-\sqrt{u}\,)^2/2.
	\label{eq: beta u limitation}
\end{equation}
It is easy\footnote{This is because direct calculation shows that the derivative of $u(1+ \ln (\beta/u))/ (1-\sqrt{u}\,)^2$ with respect to $u$ is equal to $(\sqrt{u}+\ln(\beta/u))/(1-\sqrt{u}\,)^3$, which is strictly positive for $\beta\ge 1$ and $0 < u < 1$.} to see that for each $\beta\ge 1$, the function $u(1+\ln ({\beta}/{u})) / (1-\sqrt{u}\,)^2$ is strictly increasing with respect to $u$ over $u\in(0,1)$. Hence, if (\ref{eq: beta u limitation}) holds for a $u=u_0$, it holds also for every $u\le u_0$. Therefore, (\ref{eq: beta u limitation}) puts a limit on the maximum of the sparsity for which (\ref{eq: prob bound exponent}) holds. 

Moreover, if, as done in (\ref{eq: Th prob bound temp}), we choose $r_1=r_2=c\, (1-\sqrt{u}\,)$, where $0<c<1$, then (\ref{eq: uv limitation}) states that
\begin{equation}
	u (1+\ln \frac{\beta}{u}) < c^2 (1-\sqrt{u}\,)^2/2.
	\label{eq: beta u limitation c}
\end{equation}
Similarly, this equation puts a limit on the maximum of sparsity, and since $u(1+ \ln (\beta/u))/ (1-\sqrt{u}\,)^2$ is increasing, for smaller values of $c$, this maximum on sparsity is more restricted.

\begin{figure}[tb]
\centering
\includegraphics[width=7cm]{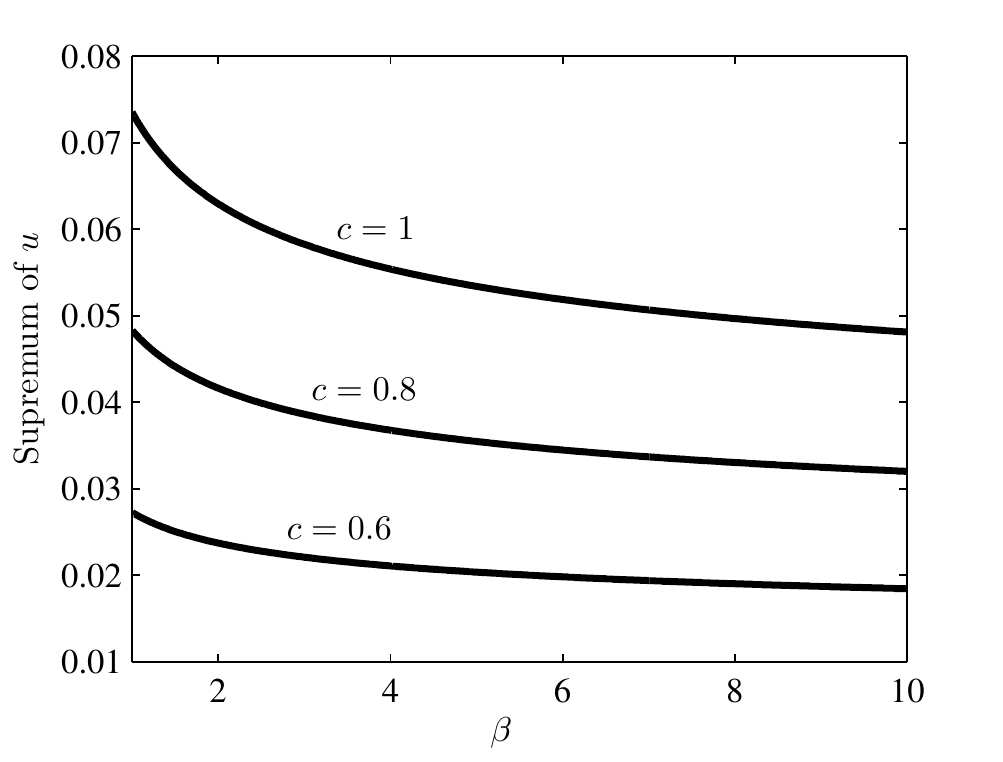}
\caption{The supremum of the values of sparsity ($u\triangleq 2p/n$) versus $\beta\triangleq m/n$ for which Theorem~\ref{th: random large dic} holds.}
\label{fig: supremum sparsity vs beta}
\end{figure}

By replacing the inequality in (\ref{eq: beta u limitation c}) with equality and solving it with respect to $u$, for each $\beta$ we obtain the supremum of $u$ for which Theorem~\ref{th: random large dic} is applicable. Figure~\ref{fig: supremum sparsity vs beta} shows the plot of this supremum versus $\beta$ for different values of $c$. Note that the value $c=1$ cannot be used, because it turns $\gamma_{r_1r_2}[2p]$ and hence the right hand side of (\ref{eq: prob bound exponent}) to infinity. It has been plotted, however, because it indicates the supremum value of sparsity for which one can choose a value for $r_2$ such that Theorem~\ref{th: random large dic} is applicable\footnote{One may note some kind of tradeoff here. Smaller $c$ results in less deviation of $\gamma_{r_1r_2}[2p]$ from $\gamma[2p]$, and hence a better upper bound in (\ref{eq: prob bound exponent}), but it decreases the sparsity for which Theorem~\ref{th: random large dic} is applicable.}. It is seen that the range of sparsity for which we can use this theorem is highly more restricted compared to the uniqueness condition $u\le 1$.

\subsection{Proofs}
We need first the following proposition that states probabilistic upper bounds on quantities $\eta_{r_1r_2}[j]$ and $\gamma_{r_1r_2}[j]$.

\begin{prop}
  \label{prop: probabilistic bound on eta gamma}
  Let $\Ab$ be an $n \times m$, $m>n$, random matrix with iid and zero-mean Gaussian entries. 
Then for each $j=1,\dots,n-1$ and for all $r_1>0$ and $0<r_2<1-\sqrt{j/n}$ we have
  \begin{equation}
     \Prob \big\{ \eta_j (\Ab) > \eta_{r_1r_2}[j]
     \big\} \le \binom{m}{j} \left( e^{-n r_1^2/2} + e^{-n r_2^2/2}
     \right),
     \label{eq: lem eta_j bound}
  \end{equation}
  and hence
  \begin{equation}
     \Prob \big\{ \gamma_j (\Ab) > \gamma_{r_1r_2}[j]
     \big\} \le \binom{m}{j} \left( e^{-n r_1^2/2} + e^{-n r_2^2/2}
     \right).
     \label{eq: lem gamma_j bound}
  \end{equation}
\end{prop}

\smallskip

\noindent {\bf Remark. } 
Any upper bound on $\binom{m}{j}$ can be used to replace this term in (\ref{eq: lem gamma_j bound}). For example~\cite[Sec. IV-A]{CandT06},
\begin{equation}
   \binom{m}{j} \le e^{m \cdot H(j/m)} \le e^{j \ln(m/j) + j},
   \label{eq: binom max}
\end{equation}
where $\forall x\in (0,1), H(x)\triangleq -x \ln x - (1-x) \ln (1-x)$.

\medskip

\begin{IEEEproof}[Proof of Proposition~\ref{prop: probabilistic bound on eta gamma}]
There is no assumption in the lemma about the variance of the entries of $\Ab$. However, since multiplying each entry of $\Ab$ by a constant does not change $\eta_j(\Ab)$ and $\gamma_j(\Ab)$, it can be assumed, without loss of generality, that this variance is equal to $\frac{1}{n}$. Let now $\Bb$ be a submatrix of $\Ab$ obtained by taking $j$ {\em `fixed'} columns of $\Ab$, and define $\eta_\Bb\triangleq \sigma_{\max}(\Bbbar)/\sigma_{\min}(\Bb)$%
. Then, from Davidson and Szarek inequalities, we have
\begin{gather}
 \Prob \left\{ \sigma_{\max}(\Bbbar) > 1 + \sqrt{\frac{m-j}{n}} + r_1 \right\} \le e^{-n r_1^2/2}, \label{eq: Abar1 bound sigmax} \\
\Prob \left\{ \sigma_{\min}(\Bb) < 1 - \sqrt{\frac{j}{n}} - r_2 \right\} \le e^{-n r_2^2/2}, \,\,\, \label{eq: A1 bound sigmin}
\end{gather}
and hence
\ifonecol
\begin{align}
  \Prob \la \eta_\Bb  > \eta_{r_1r_2}[j]\ra & \le 
  \Prob \la \sigma_{\max}(\Bbbar)>1 + \sqrt{\frac{m-j}{n}} + r_1 \ra 
   + \Prob \la \sigma_{\min}(\Bb)<1 - \sqrt{\frac{j}{n}} - r_2 \ra \nonumber \\
  & \le e^{-n r_1^2/2} + e^{-n r_2^2/2}.
 \label{eq: lem eta bound temp1}
\end{align}
\else
\begin{align}
  \Prob \la \eta_\Bb  > \eta_{r_1r_2}[j]\ra & \le 
  \Prob \la \sigma_{\max}(\Bbbar)>1 + \sqrt{\frac{m-j}{n}} + r_1 \ra  \nonumber \\
  & + \Prob \la \sigma_{\min}(\Bb)<1 - \sqrt{\frac{j}{n}} - r_2 \ra \nonumber \\
  & \le e^{-n r_1^2/2} + e^{-n r_2^2/2}.
 \label{eq: lem eta bound temp1}
\end{align}
\fi
$\eta_j(\Ab)$ is the maximum of $\eta_\Bb$ on all $\binom{m}{j}$ possible choices for $\Bb$. Therefore, using the union bound,
\ifonecol
\begin{equation*}
  \Prob \la \eta_j(\Ab)  > \eta_{r_1r_2}[j]\ra  \le 
  \Prob \la \bigcup_{\Bb\in {\cal P}_j(\Ab)} \eta_\Bb  > \eta_{r_1r_2}[j] \ra 
  \le \binom{m}{j} \left(e^{-n r_1^2/2} + e^{-n r_2^2/2}\right),
\end{equation*}
\else
\begin{equation*}
 \begin{split}
  \Prob \la \eta_j(\Ab)  > \eta_{r_1r_2}[j]\ra & \le 
  \Prob \la \bigcup_{\Bb\in {\cal P}_j(\Ab)} \eta_\Bb  > \eta_{r_1r_2}[j] \ra \\
  & \le \binom{m}{j} \left(e^{-n r_1^2/2} + e^{-n r_2^2/2}\right),
 \end{split}
\end{equation*}
\fi
which completes the proof of (\ref{eq: lem eta_j bound}). 
We have also (\ref{eq: lem gamma_j bound}), because the events $\la\eta_j(\Ab) > \eta_{r_1r_2}[j]\ra$ and $\la\gamma_j(\Ab) > \gamma_{r_1r_2}[j]\ra$ are identical (while $1-\sqrt{j/n}-r_2>0$).
\end{IEEEproof}

\medskip

\begin{IEEEproof}[Proof of Theorem~\ref{th: Prob bound}]
From (\ref{eq: tight bound}), if $\bar\gamma'(\Ab) \le \gamma_{r_1r_2}[\ell]$ then $\nt{\hsbb-\sbb_0} \le (\gamma_{r_1r_2}[\ell]) \alpha_{\hsbb,\ell}$. This is equivalent to say that if $\nt{\hsbb-\sbb_0} > (\gamma_{r_1r_2}[\ell]) \alpha_{\hsbb,\ell}$ then $\bar\gamma'(\Ab) > \gamma_{r_1r_2}[\ell]$. Therefore 
  \begin{equation}
    \Prob \big\{ \nt{\hsbb-\sbb_0}>(\gamma_{r_1r_2}[\ell]) \alpha_{\hsbb,\ell} \big\} \le
    \Prob \la \bar\gamma'(\Ab) > \gamma_{r_1r_2}[\ell]\ra.
    \label{eq: ThmProbBound temp1}
  \end{equation}
By Lemma~\ref{lem: g increasing}, $\sqrt{m}<\gamma_{r_1r_2}[\ell]$ and hence $\bar\gamma'(\Ab) > \gamma_{r_1r_2}[\ell]$ is equivalent to $\max_{1\le j \le \ell} \gamma_j(\Ab)>\gamma_{r_1r_2}[\ell]$. Therefore, from the union bound,
\ifonecol
\begin{equation}
    \Prob \la \bar\gamma'(\Ab) > \gamma_{r_1r_2}[\ell]\ra  
   \le \sum_{i=1}^\ell \Prob \Big\{ \gamma_j(\Ab)>\gamma_{r_1r_2}[\ell] \Big\}
     \le \sum_{i=1}^\ell \Prob \Big\{ \gamma_j(\Ab)>\gamma_{r_1r_2}[j] \Big\} 
     \label{eq: ThmProbBound temp2}
\end{equation}
\else
\begin{align}
    \Prob &\la \bar\gamma'(\Ab) > \gamma_{r_1r_2}[\ell]\ra  
   \le \sum_{i=1}^\ell \Prob \Big\{ \gamma_j(\Ab)>\gamma_{r_1r_2}[\ell] \Big\} \nonumber \\
    & \le \sum_{i=1}^\ell \Prob \Big\{ \gamma_j(\Ab)>\gamma_{r_1r_2}[j] \Big\}, \label{eq: ThmProbBound temp2}
\end{align}
\fi
where in the last inequality, Lemma~\ref{lem: g increasing} has been used. Now, combining (\ref{eq: ThmProbBound temp1}), (\ref{eq: ThmProbBound temp2}) and (\ref{eq: lem gamma_j bound}) proves the theorem.
\end{IEEEproof}

\medskip

\begin{IEEEproof}[Proof of Theorem~\ref{th: random large dic}]
Let $r\triangleq\min(r_1,r_2)$ and $P\triangleq\Prob \big\{ \nt{\hsbb-\sbb_0}>(\gamma_{r_1r_2}[2p]) \alpha_{\hsbb,2p} \big\}$. Then, from (\ref{eq: Th prob bound p}) we have
\begin{equation}
    P<2p \binom{m}{2p} \left( e^{-n r_1^2/2} + e^{-n r_2^2/2} \right) 
     <4p \binom{m}{2p} e^{-n r^2/2}.
\end{equation}
Hence, from (\ref{eq: binom max}),
\begin{equation}
  \begin{split}
    P &<4p \cdot \exp\left\{2p\ln\frac{m}{2p}+2p-\frac{n r^2}{2}\right\} \\
      & = 4p\cdot  \exp\left\{-n \left[\frac{r^2}{2} - \frac{2p}{n}\left(1+\ln\frac{m}{2p}\right) \right]\right\}\cdot
  \end{split}
\end{equation}
When $n$ grows to infinity, the coefficient of $-n$ in the exponent converges to the constant $r^2/2-u (1+\ln v)$, which is positive by the assumption (\ref{eq: uv limitation}), and hence, $P$ is upper bounded by an exponentially decreasing function.
\end{IEEEproof}

\section{Conclusion}
In this paper, we studied upper bounds for the estimation error $\nt{\hsbb-\sbb_0}$. We saw that such bounds can be constructed only based on $\hsbb$, and without knowing $\sbb_0$ (the existence of a sparse $\sbb_0$ satisfying $\nz{\sbb_0}<\frac{1}{2}\spark(\Ab)$ has been assumed). We have also presented a tight upper bound for this error. Besides being tight, this bound does not impose any assumption on the normalization of the atoms of the dictionary, which enabled us to study random dictionaries (which are used \eg\ in compressed sensing).

As a result, our bounds guaranty that whenever $\hsbb$ is only approximately (not exactly) sparse, it would be not too far from $\sbb_0$, and the upper bound on their distance is determined by the properties of the dictionary ($\Ab$). This upper bound decreases also when $\hsbb$ is sparse with a better approximation. In this point of view, our bounds can be seen as a generalization of the uniqueness theorem to the case $\hsbb$ is only approximately sparse. Moreover, these bounds show that whenever $\nz{\sbb_0}$ grows, to obtain a predetermined guaranty on the maximum of $\nt{\hsbb-\sbb_0}$, $\hsbb$ is needed to be sparse with a better approximation. This can be seen as an explanation to the fact that the estimation quality of sparse recovery algorithms degrades whenever $\nz{\sbb_0}$ grows.

 We also studied the noisy case, and we saw that constructing a general upper bound for this case is not easy. Hence, we did not study random dictionaries for this noisy case, which can be a subject for future investigations.

\section{Appendix}

\subsection{Proof of Lemma~\ref{lem: gamma1 lt gamma0}} \label{app: gamma1 lt gamma0}
We will need the following lemma:
\begin{lemma}\label{lem: min sig max}
Let $\Bb$ be an $n \times p$ matrix, $p \ge n$, with unit $\ell^2$ norm columns. Then $\sigma_{\max}(\Bb)\ge \sqrt{p/n}\ge 1$.
\end{lemma}
\begin{IEEEproof}
The singular values of $\Bb$ are the square root of eigenvalues of $\Cb_{p \times p}\triangleq \Bb^T \Bb$. Moreover, since the columns of $\Bb$ have unit Euclidean norms, the main diagonal elements of $\Cb$ are all equal to 1. Therefore,  $\sum_{i=1}^p \lambda_i(\Cb)=\mbox{tr}(\Bb)=p$, where $\lambda_i(\Cb)$ denote the eigenvalues of $\Cb$. On the other hand, the rank of $\Cb$ is at most $n$, and hence there are at most $n$ nonzero $\lambda_i$'s. Therefore
\begin{equation*}
	p = \sum_{i=1}^p \lambda_i(\Cb) \le n \lambda_{\max}(\Cb) \Rightarrow \lambda_{\max}(\Cb) \ge \frac{p}{n},
\end{equation*}
which completes the proof.
\end{IEEEproof}

\begin{IEEEproof}[Proof of Lemma~\ref{lem: gamma1 lt gamma0}]
  From the definition (\ref{eq: eta def}), for $j=1$, $\Ab_1$ has only one column and hence $\sigma_{\min}(\Ab_1)=1$ using Lemma~\ref{lem: sv of 1 and 2 col}. Moreover, $\Abbar_1$ is an $n \times (m-1)$ matrix. We write
\begin{gather*}
\gamma_1(\Ab) \ge \sqrt{m} \Leftrightarrow
\sqrt{(m-1)\left( 1 + \sigma_{\max}^2(\Abbar_1)\right)} \ge \sqrt{m} \\ \Leftrightarrow
1 + \sigma_{\max}^2(\Abbar_1) \ge \frac{m}{m-1}=1+\frac{1}{m-1} \\
\Leftrightarrow 
\sigma_{\max}^2(\Abbar_1) \ge \frac{1}{m-1},
\end{gather*}
which holds by Lemma~\ref{lem: min sig max}, because $\sigma_{\max}^2(\Abbar_1) \ge 1 \ge \frac{1}{m-1} $. 
\end{IEEEproof}

\subsection{Proof of Lemma~\ref{lem: g increasing}}
To prove that $\gamma_{r_1r_2}[j]$ is strictly increasing with respect to $j$, we state the following lemma, in which, we first define a function $\Gamma(x)$, $x \in \Rr$, such that $\gamma_{r_1r_2}[j]$'s are scaled samples of this function (more precisely $\gamma_{r_1r_2}[j]=\sqrt{n}\,\Gamma(j/n)$) for appropriate values of the parameters of the function. Then, we show that $\Gamma(x)$ is itself strictly increasing, and hence so are its samples.
\begin{lemma}\label{lem: analog g increasing}
Let $p,a,b$ be real numbers with $a\ge 0$, $p\ge b^2 >0$. Then the function $\Gamma(\cdot)$, defined below, is strictly increasing on the interval $[0, b^2)$:
\begin{equation}
	\Gamma(x) \triangleq  \sqrt{(p-x) \left[ 
	1 + \left(
	   \frac{a+\sqrt{p-x}}{b-\sqrt{x}}
	\right)^2
	\right]}\cdot
  \label{eq: g def}
\end{equation}
\end{lemma}

\medskip

Before going to the proof, note that $\gamma_{r_1r_2}[j]=\sqrt{n}\,\Gamma(j/n)$ for $p=m/n$, $a=1+r_1$ and $b=1-r_2$.

\medskip

\begin{IEEEproof}[Proof of Lemma~\ref{lem: analog g increasing}]
We have to prove that $g(x)\triangleq \Gamma^2(x)$ is increasing on $[0,b^2)$, and hence
we have to prove that $g'(x)>0, \forall x\in[0,b^2)$. 
By defining	$h(x)\triangleq (\sqrt{p-x} \,) \frac{a+\sqrt{p-x}}{b-\sqrt{x}}=\frac{a\sqrt{p-x}+(p-x)}{b-\sqrt{x}}$
we have $g(x) = (p-x)+ h^2(x)$, and hence $g'(x) = -1 + 2h(x)h'(x)$.
Consequently, we have to prove $2h(x)h'(x)>1, \forall x\in[0,b^2)$.
Direct calculations show that $2h(x)h'(x)$ is equal to
\begin{equation*}
	\frac{(a+\spx)\left[
	-(a+2\spx)(\bsx)+ \frac{p-x}{\sqrt{x}}(a+\spx)
	\right]}{(\bsx)^3},
\end{equation*}
and hence $2h(x)h'(x)>1$ is equivalent to
\ifonecol
\begin{equation}
	\frac{(p-x)(a+\spx)^2}{\sqrt{x}} > (\bsx)^3 + (a+\spx)(a+2\spx)(\bsx).
	\label{eq: To prove}
\end{equation}
\else
\begin{multline}
	\frac{(p-x)(a+\spx)^2}{\sqrt{x}} > (\bsx)^3 + \\ +(a+\spx)(a+2\spx)(\bsx).
	\label{eq: To prove}
\end{multline}
\fi
To prove (\ref{eq: To prove}) we multiply both sides by $\sqrt{x}/[(a+\spx)^2(\bsx)]$ and write it as
\begin{equation}
	\frac{p-x}{\bsx} > \sqrt{x} \left[
	  \left(
	    \frac{\bsx}{a+\spx}
	  \right)^2 + 1 +
	  \frac{\spx}{a+\spx}
	\right].
	\label{eq: temp1 ineq}
\end{equation}
Note that from $p\ge b^2$ we have
\begin{equation}
	\frac{p-x}{\bsx} \ge \frac{b^2-x}{\bsx}=b+\sqrt{x},
\end{equation}
and hence to prove (\ref{eq: temp1 ineq}) it is sufficient to prove that
\begin{equation}
	 b+\sqrt{x} > \sqrt{x} \left[
	  \left(
	    \frac{\bsx}{a+\spx}
	  \right)^2 + 1 + 
	  \frac{\spx}{a+\spx}
	\right],
\end{equation}
which by multiplying both sides by $(a+\spx)^2$ is equivalent to
\begin{equation}
	 b (a+\spx)^2 - \sqrt{x} \left[
	  \left(
	    {\bsx}
	  \right)^2 + 
	  \spx\,(a+\spx)
	\right] >0.
	\label{eq: temp 2 ineq}
\end{equation}
Doing some algebraic manipulations, the left hand side of the above inequality is equal to
\begin{equation}
	a^2 b + ab\spx + a\spx(\bsx)+(\bsx)(p-b\sqrt{x}),
\end{equation}
and hence (\ref{eq: temp 2 ineq}) holds because the first 3 terms of the above expression are nonnegative (note that $a$ may be equal to zero), and the last term is positive from $b>\sqrt{x}$ and $p>b\sqrt{x}$ (because $p\ge b^2 =b.b>b\sqrt{x}$). 
\end{IEEEproof}

\medskip

\begin{IEEEproof}[Proof of Lemma~\ref{lem: g increasing}]
  Note that $\gamma_{r_1r_2}[j]=\sqrt{n}\,\Gamma(j/n)$ for $p=m/n$, $a=1+r_1$ and $b=1-r_2$, where $\Gamma(\cdot)$ is as defined in (\ref{eq: g def}). Now, since $p=\frac{m}{n} > 1 \ge (1-r_2)^2=b^2>0$, the conditions of Lemma~\ref{lem: analog g increasing} are satisfied, and hence that lemma insures that $\gamma_{r_1r_2}[j]$ is strictly increasing. 

  To prove $\gamma_{r_1r_2}[j] > \sqrt{m}$, we note that it is equivalent to
  \begin{equation*}
     (m-1)(1+\eta^2_{r_1r_2}[j]) > m \Leftrightarrow \eta^2_{r_1r_2}[j] > \frac{1}{m-1}\cdot
  \end{equation*}
  which holds because $\eta^2_{r_1r_2}[j]>1$ and $1\ge\frac{1}{m-1}$.
\end{IEEEproof}

\begin{figure}
  \unitlength 6.5mm
  \centering
  \begin{picture}(9.5, 6)(-0.5, -0.5)
  \thinlines

  \put(-0.5,0){\vector(1,0){9}}
  \put(0,-0.5){\vector(0,1){5.5}}

  \put(7,-0.1){\line(0,1){0.2}}

  \multiput(6.6,-0.1)(0,0.6){9}{\line(0,1){0.4}}

  \put(4,0.5){$\dots$}

  \put(8,-1.25){\makebox(1,1)[tl]{$x$}}
  \put(0.5,-1.2){\makebox(1,1)[tc]{\small$\frac{1}{n}$}}
  \put(1.5,-1.2){\makebox(1,1)[tc]{\small$\frac{2}{n}$}}
  \put(2.5,-1.2){\makebox(1,1)[tc]{\small$\frac{3}{n}$}}
  \put(5.5,-1.2){\makebox(1,1)[tc]{\small$\frac{n-1}{n}$}}
  \put(6.5,-1.2){\makebox(1,1)[tc]{\small$1$}}

  \put(6.65,0.05){\vector(1,1){0.8}}
  \put(7.45,.75){\makebox(1,1)[bl]{\tiny$(1-r_2)^2$}}

  \put(4.8,1.6){\vector(-1,1){1}}
  \put(2.8,2.6){\makebox(1,1)[br]{\footnotesize$\sqrt{n}\,\Gamma(x)$}}

  \thicklines
  \qbezier(0,0.5)(6.5,0.6)(6.5,5)

  \put(1,0){\line(0,1){0.55}}
  \put(1,0.55){\circle*{0.2}}
  \put(0.5,0.8){\makebox(1,1)[bc]{\footnotesize$\Gamma_1$}}

  \put(2,0){\line(0,1){0.64}}
  \put(2,0.64){\circle*{0.2}}
  \put(1.5,0.89){\makebox(1,1)[bc]{\footnotesize$\Gamma_2$}}

  \put(3,0){\line(0,1){0.85}}
  \put(3,0.85){\circle*{0.2}}
  \put(2.5,1.1){\makebox(1,1)[bc]{\footnotesize$\Gamma_3$}}

  \put(6,0){\line(0,1){2.9}}
  \put(6,2.9){\circle*{0.2}}
  \put(4.9,3.1){\makebox(1,1)[bc]{\footnotesize$\Gamma_{n-1}$}}

  \end{picture}
  \caption{A typical graph of $\Gamma(x)$ defined in (\ref{eq: g def}) and the values $\gamma_{r_1r_2}[j]$ defined in (\ref{eq: g r1r2 def}). Note: $\Gamma_j$ in the figure stands for $\gamma_{r_1r_2}[j]=\sqrt{n}\, \Gamma(j/n)$.}
  \label{fig: g g_r1r2}
\end{figure}
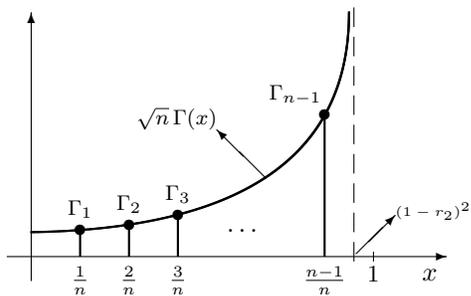

Figure~\ref{fig: g g_r1r2} shows a typical graph of ${\Gamma(x)}$ and $\gamma_{r_1r_2}[j]$ (denoted by $\Gamma_j$ in the figure).


\begin{IEEEbiography}[{\includegraphics[width=1in,height=1.25in,clip,keepaspectratio]{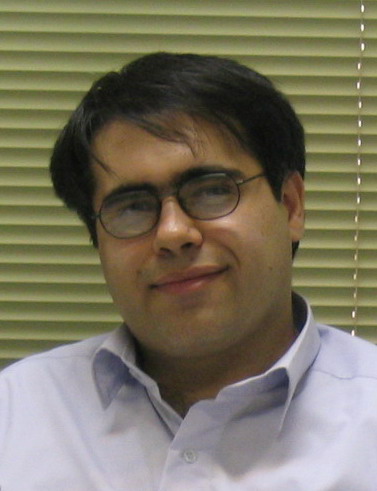}}]%
{{\bf Massoud Babaie-Zadeh} \rm (M'04-SM'09) received the B.S. degree in electrical engineering from Isfahan University of Technology, Isfahan, Iran in 1994, and the M.S degree in electrical engineering from Sharif University of Technology, Tehran, Iran, in 1996, and the Ph.D degree in Signal Processing from Institute National Polytechnique of Grenoble (INPG), Grenoble, France, in 2002.\\
\hspace*{1.5em}Since 2003, he has been a faculty member of the Electrical Engineering Department of Sharif University of Technology, Tehran, IRAN, firstly as an assistant professor and since 2008 as an associate professor. 
He was also an invited professor at the INPG, Grenoble, France and at the University of Evry-Val-d'Essonne, Evry, France, in summers 2006 and 2008, respectively. Furthermore, he was a visiting professor at the University of Minnesota from October 2010 to September 2011.
His main research areas are Blind Source Separation (BSS) and Independent Component Analysis (ICA), Sparse Signal Processing, and Statistical Signal Processing.\\
\hspace*{1.5em}Dr. Babaie-Zadeh received the best Ph.D. thesis award of INPG for his Ph.D. dissertation.
}
\end{IEEEbiography}

\begin{IEEEbiography}[{\includegraphics[width=1in,height=1.25in,clip,keepaspectratio]{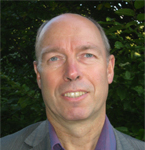}}]%
{{\bf Christian Jutten} \rm (AM'92-M'03-SM'06-F'08) received the PhD degree in 1981 and the Docteur ès Sciences degree in 1987 from the Institut National Polytechnique of Grenoble (France). After being associate professor in the Electrical Engineering Department (1982-1989) and visiting professor in Swiss Federal Polytechnic Institute in Lausanne (1989), he became full professor in University Joseph Fourier of Grenoble, more precisely in the sciences and technologies department. For 30 years, his research interests are learning in neural networks, blind source separation and independent component analysis, including theoretical aspects (separability, source separation in nonlinear mixtures) and applications (biomedical, seismic, speech).  He is author or co-author of more than 55 papers in international journals, 4 books, 18 invited papers and 160 communications in international conferences. He was co-organizer of the 1st International Conference on Blind Signal Separation and Independent Component Analysis (Aussois, France, January 1999). He has been a scientific advisor for signal and images processing at the French Ministry of Research (1996-1998) and for the French National Research Center (2003-2006). He has been associate editor of IEEE Trans. on Circuits and Systems (1994-95). He is a member of the technical committee "Blind signal Processing" of the IEEE CAS society and of the technical committee "Machine Learning for signal Processing" of the IEEE SP society. He received the EURASIP best paper award in 1992 and Medal Blondel in 1997 from SEE (French Electrical Engineering society) for his contributions in source separation and independent component analysis, and has been elevated as a Fellow IEEE and a senior Member of Institut Universitaire de France in 2008.
}
\end{IEEEbiography}

\begin{IEEEbiography}[{\includegraphics[width=1in,height=1.25in,clip,keepaspectratio]{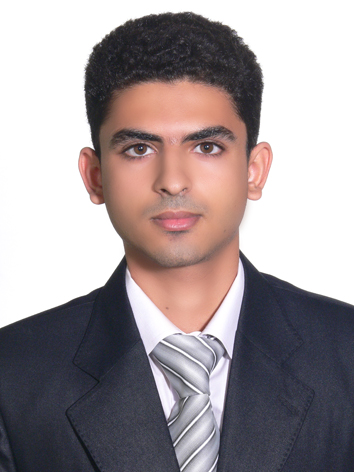}}]%
{{\bf Hosein Mohimani} \rm was born in Bushehr, Iran, in 1985. He received a double major (Electrical Engineering / Mathematics) B.S.~degree from Sharif University of Technology, Tehran, Iran in 2008. He is currently working toward his Ph.D degree in Electrical and Computer Engineering in the University of San Diego.
}
\end{IEEEbiography}

\end{document}